\documentclass[12pt,a4paper,twoside]{article}
\usepackage{amssymb}
\usepackage{graphicx}

\usepackage{setspace}

\newtheorem{conjecture}{Conjecture}

\newtheorem{definition}{Definition}

\begin{document}

\title{An efficient centralized binary multicast network coding algorithm for any cyclic network}
%
%

\author{\normalsize \'{A}ngela I. Barbero$^1$ and {\O}yvind Ytrehus$^2$\\
\small $^1$Dept. of Applied Mathematics, University of Valladolid, 47011 Valladolid, Spain.\\
\small E-mail: angbar@wmatem.eis.uva.es\\
\small $^2$Dept. of Informatics, University of Bergen, N-5020
Bergen, Norway\\
\small E-mail:oyvind@ii.uib.no }
\date{\today}
\maketitle

\begin{abstract}
\noindent We give an algorithm  for finding network encoding and
decoding equations
for error-free multicasting networks with multiple sources and
sinks. The algorithm given is efficient (polynomial complexity) and
works on any kind of network (acyclic, link cyclic, flow cyclic, or
even in the presence of knots).  The key idea will be the
appropriate use of the delay (both natural and additional) during
the encoding. The resulting code will always work with finite delay
with binary encoding coefficients.
\end{abstract}


\section{Introduction}

We consider centralized algorithms for designing the network
coding equations for multicast \cite{Als} in a network with
directed edges. Early work in this area focused on acyclic
networks (see \cite{SET}, \cite{jaggi} and the references
therein). Cyclic networks were considered in \cite{cyclopathic}
and in \cite{knotwork}, and \cite{ErezFederISIT04}. In
\cite{cyclopathic} we developed, based on the Linear Information
Flow (LIF) algorithm in \cite{SET}, the LIFE and LIFE-CYCLE
algorithms. These will find a linear encoding (if one exists) in
networks that are, respectively, link cyclic and simple flow
cyclic (see \cite{cyclopathic} and Section~\ref{sec_notation} for
definitions of different types of cyclicity).

Thus the state of the art prior to this paper is that centralized
encoding can calculate, in polynomial time, a valid encoding for
acyclic and simple cyclic networks. However, many networks occurring
in practice may, in fact, contain complex cyclic structures.

There are two contributions of this paper:
\begin{enumerate}
 \item We extend the algorithms in \cite{cyclopathic} by applying
 Mason's formula \cite{LinCostello},\cite{MasonPaperMengetal} for signal propagation in a cyclic graph. The
 resulting new algorithm (LIFE*) will calculate a valid encoding for
 arbitrary cyclic and acyclic networks. The complexity is similar to that of
 the LIFE algorithm.
 \item We propose a simple \emph{binary} encoding scheme that exploits the
 natural delay inherent in the network. This scheme is related to
 those proposed in \cite{soljanin}, \cite{ErezFederISIT04}, \cite{YeungISIT06}, and \cite{FrancisITW07}, and also to the
 randomized version of the LIF algorithm \cite{jaggi}.
\end{enumerate}

In Section~\ref{sec_notation} we give an overview of the necessary
notation and previous results. Further, we will recall the two
different notions of cyclicity and the two different types of flow
cycles. Next, in Section~\ref{sec_life+d} we present the LIFE*
algorithm. Although the algorithm works over any field, we propose
that the delay based binary encoding scheme demonstrated in
Section~\ref{sec_life+d} is particularly well suited for the LIFE*
algorithm. Finally in Section~\ref{sec_considerations} we discuss
complexity issues and practical aspects.
We have included an appendix, which makes up the bulk of the paper,
with detailed examples that show how the algorithms work on the
different kinds of networks.

\section{Notation and previous results}
\label{sec_notation}

The notations used in this paper follow and expand those used in \cite{cyclopathic}.

Consider the network $G$, where $G=(V,E)$ is a directed multigraph.
$V$ represents the set of nodes and $E$ the set of edges or links,
each of unit capacity. A pair of nodes may be connected by one or
multiple edges.
 Let $S=\{s_1,s_2,\ldots,s_h\}\subset V$ be
the set of unit rate information sources and let
$T=\{t_1,t_2,\ldots,t_r\}\subset V$ be the set of sinks. We will
assume that each source (synchronously) generates one symbol at each
(discrete) time instant. Let $x\in \mathbb{Z}$ denote the time. We
will denote by $\sigma_i(x)\in \mathbb{F}_2$ the symbol produced by
source $s_i$ at time $x$. In case the given network has one unique
source $s$ that sends information at rate $h$ we will simply create
virtual sources $s_i, \ldots, s_h$ and link them to the actual
source $s$.
Note that most real life networks can be precisely or approximately
described by a network $G$ as outlined here.

For each edge $e\in E$ we will denote $start(e)$ and $end(e)$ the
nodes at which $e$ starts and ends, respectively. As usual, a
\emph{path} from node $u$ to node $v$ of length $l$ is a sequence
$\{e_i \in E: i = 1,\ldots, l\}$ such that $u = start(e_1)$, $v
=end(e_l)$, and  $start (e_{i+1}) = end(e_i)$, for $i=1,\ldots,l-1$.

The complete set of symbols $\sigma_i(x), \;i=1\ldots h$ will be
called generation $x$. The object of the algorithm is to find an assignment of equations
in such a way that each sink
$t\in T$ at each time $x+d_t$ can complete the decoding of the whole
generation $x$, where $d_t$ is a (finite) constant for $t$ denoting the total
delay associated to that sink. It is assumed that $\sigma_i(x)=0,
i=1\ldots h$ for any $x<0$.


 We assume that the transmission of a
symbol on each link $e$ has a unit delay associated with it, that is
to say, if a symbol $s$ is being carried by edge $e$ at time $x$,
that symbol, once processed at node $end(e)$, can be carried by an
edge $e'$ at time $x+1$, for any edge $e'$ with $start(e')=end(e)$.
Apart from this intrinsic delay, as we will see, the encoding
process might assign extra delays in order to satisfy certain
required conditions.

Let $D$ denote the linear delay operator. The way in which the operator
works is as follows:

\[D(\sigma(x))=D\sigma(x)=\sigma(x-1)\]

and it is extended by linearity.

\[D(\sigma(x)+\xi(x))=D\sigma(x)+D\xi(x)=\sigma(x-1)+\xi(x-1)\]
\[D^i\sigma(x)=D(D^{i-1}\sigma(x))=\sigma(x-i)\; \forall i\in{ \mathbb Z}_0^+\]
\[(D^i+D^j)\sigma(x)=D^i\sigma(x)+D^j\sigma(x)=\sigma(x-i)+\sigma(x-j)\]

Also, abusing notation, we extend the operator in order to work with
negative exponents in the following way.

\[D^{-1}\sigma(x)=\sigma(x+1)\]

A \emph{flow path} $f^{s,t}$ is simply a path from a source $s$ to a
sink $t$. We will assume that for each sink $t \in T$ there exists
an edge disjoint set of flow paths  $f^t = \{f^{s_i,t}, i=1,\ldots
h\}$. Following the notation of \cite{SET} we will call $f^t$ the
\emph{flow} for sink $t$. The minimal subgraph of $G$ that contains
all flows (and their associated nodes) will be called the \emph{flow
path graph}, and can be determined from $G$ by a suitable polynomial
algorithm.
>From the perspective of encoding as discussed in this paper, we
ignore the issue of determining the flow path graph and the flows,
and assume that the network is a flow path graph, that the flows are
known, and that every edge is on some flow path.


The results in \cite{Als} guarantee that in such a network, all
the sinks can receive all the $h$ input symbols produced by the
$h$ sources, and the results in \cite{Li} state that it can be
done using linear coding on the network. In this paper we deal
with linear coding, so each edge will be encoded using a linear
encoding equation.


\begin{definition}
A \emph{link cyclic} network is a network where there exists a
cyclic subset of edges, i. e., a set
$\{e_1, e_2,\ldots,e_{k},e_{k+1}=e_1\} \subset E$ for some positive
integer $k$ such that $end(e_i) = start(e_{i+1})$ for $1 \leq i <
k$. The set of edges $\{e_1,e_2,\ldots,e_{k},e_{k+1}=e_1\}$ is a \emph{link cycle}.
If no such cycle exists, the network is \emph{link acyclic} or
simply \emph{acyclic}.
\end{definition}

Suppose $e$ is an edge that lies on the flow path $f^{s_i,t_j}$. We
will denote by $f_{\leftarrow}^{t_j}(e)$ the predecessor of edge $e$
in that path. There is no ambiguity in the notation: $e$ can lie on
several flow paths $f^{s_i,t_j}$ for different $t_j$, but not for
different $s_i$ and the same $t_j$, since all the flow paths
arriving in $t_j$ are edge disjoint. Thus, once $t_j$ is fixed, $e$
can only have one predecessor in the flow to $t_j$, that is,
$f_{\leftarrow}^{t_j}(e)$ is the predecessor of $e$ in the only flow
path arriving in $t_j$ that contains $e$. In the same way we denote
by $f_{\rightarrow}^{t_j}(e)$ the successor of $e$ in that flow
path.

Let $T(e)\subseteq T$ denote the set of sinks $t$ that use $e$ in
some flow path $f^t$, and let $P(e)=\{f_{\leftarrow}^{t}(e)\;
|\;t\in T(e)\}$ denote the set of all predecessors of edge $e$.
Note that all edges will have some predecessor ($P(e)\neq
\emptyset$), except those with $start(e)=s_i$ for $i\in\{1,\ldots,
h\}$.

We will introduce some extra notation to denote the temporal order
induced in the edges by the flow paths in $f$. When two edges $e_1$
and $e_2$ lie on the same flow path and $e_1$ is the predecessor of
$e_2$ in that path, we will write $e_1\prec e_2$. We will use
transitivity to define relationships among other pairs of edges that
lie on the same path but are not consecutive. We observe that in
each path the relation $\prec$ defines a total order in the edges
that form that path, since a path that contains cycles, that is,
edges that satisfy $e_1\prec e_2\prec \cdots \prec e_n \prec e_1$
can be simplified by avoiding taking the trip around that cycle.

\begin{definition}
A \emph{flow acyclic} network is a network where the relation
$\prec$ defines a partial ordering in $E$. If the relation does
not define a partial ordering, the network is flow cyclic.
\end{definition}

Note that a flow cyclic network is always link cyclic, but the
converse is not true.

\begin{definition} A \emph{simple flow cycle} is a link cycle $\{e_1,e_2, \ldots,e_k,e_{k+1}=e_1\}$
such that for each $i=1,\ldots ,k$ there exists a flow path
$f^{s_i,t_i}$ that implies $e_i\prec e_{i+1}$. We observe here that
must be at least two distinct flow paths traversing the cycle.
\end{definition}

\begin{definition}
 A \emph{flow knot} or simply a \emph{knot} is
formed by two or more simple flow cycles that share one or more
edges.
\end{definition}

Figure~1 shows an example of a knot. This particular knot forms part
of the network in Example 4 in Figure~2.

\begin{figure}[htbp]
\centerline{\includegraphics[width=7cm,keepaspectratio=true,angle=0]{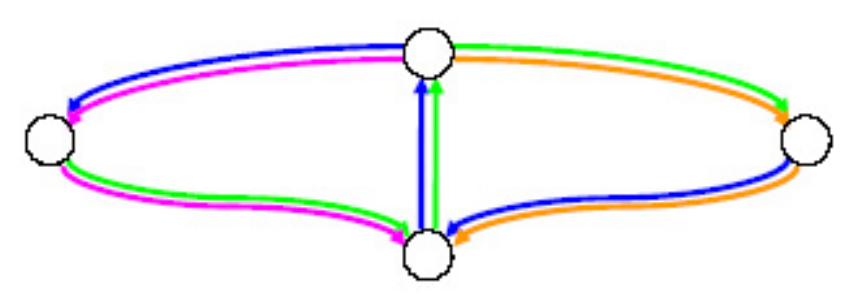}}
\label{Fig_Knot} \caption{A flow knot.}
\end{figure}

We illustrate these concepts with some examples, shown in Figure~2,
in which we show the flow path graphs for some networks. To make it
easier to see, each sink is represented with a different color and
all the flow paths arriving in that sink are drawn in the same
color. One can easily check that the first example is an acyclic
network (it is one of the so-called combination networks), the
second is link cyclic but flow acyclic while Examples 3 and 4 are
flow cyclic. Example 3 is the same cyclic network presented in
\cite{Als}, while Example 4 has been created by us to illustrate a
network with a flow knot.

\begin{figure}[htbp]
\centerline{\includegraphics[width=15cm,keepaspectratio=true,angle=0]{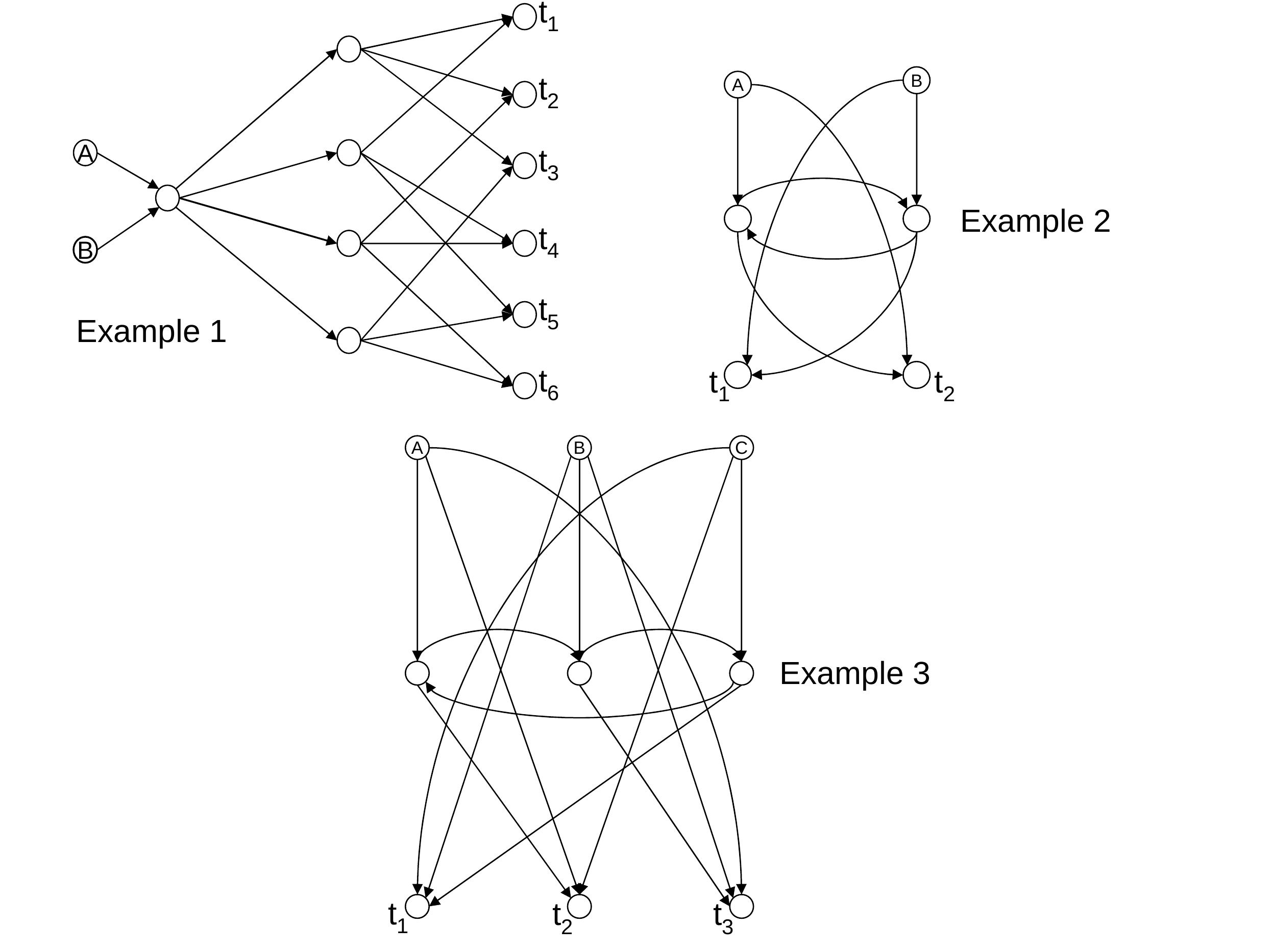}}
\centerline{\includegraphics[width=15cm,keepaspectratio=true,angle=0]{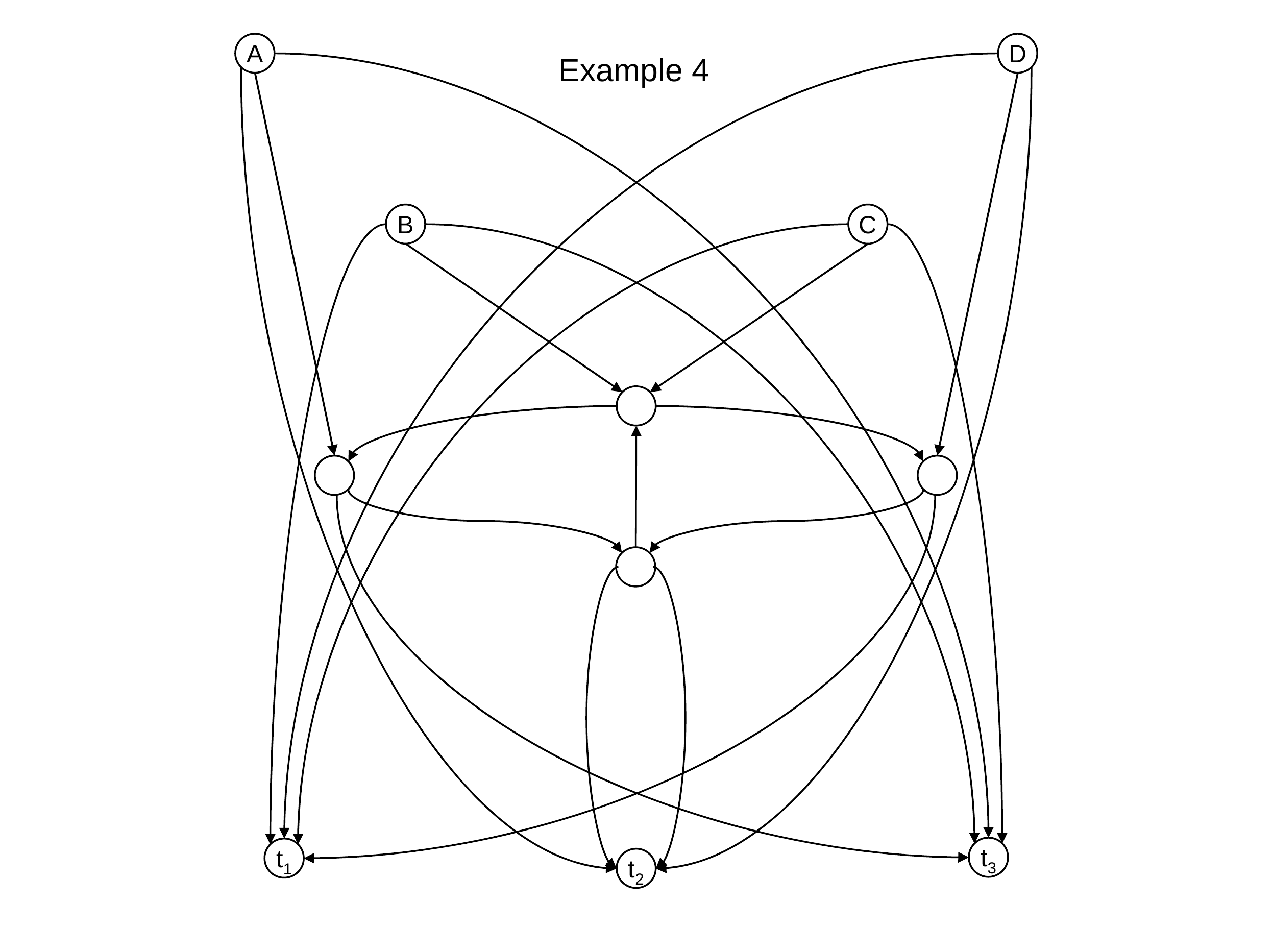}}
\label{Fig_Examples} \caption{Examples of different types of
networks.}
\end{figure}

\section{ The LIFE* algorithm}
\label{sec_life+d}

In the whole process of encoding we will observe two basic
principles.

\begin{itemize}

\item {\bf Pass on information (PI) principle:} \cite{cyclopathic} When
encoding each edge $e\in E$, all the symbols carried by the incoming
edges $f_{\leftarrow}^{t}(e), t\in T(e)$ will contribute.
This is because assigning coefficient 0 to any of those predecessors
means that the flow carried on that predecessor is stopped at that
point, meaning in turn that the actual flow path graph used is
different from the one established initially. \footnote{Observe that
this principle makes sense if and only if the flow path graph is
``reasonably efficient''. It may require some effort to calculate a
reasonably efficient flow path graph, but if that work has been done
it is wise, in terms of complexity, to rely on the provided flow
path graph. The connection between flow path graph calculation and
network coding can offer interesting complexity trade-offs, but this
discussion is beyond the scope of this paper.}

\item {\bf Old symbol removal (OSR) principle:} \cite{cyclopathic} To avoid symbols circulating
endlessly on each flow cycle, they should be removed at the entrance
point by the node through which they entered the cycle. We will
explain this in more detail in \ref{cyclic part}.

\end{itemize}

In what follows we will consider two ways of expressing the symbol
$v_e(x)\in\mathbb{F}_2$ to be transmitted on edge $e$ at time $x$.

The \emph{local encoding equation} specifies the action of each
node. The general local encoding equation is

\begin{equation} \label{eq_localencoding0}
v_e(x)=\sum_{p\in P(e)}\pi(p,e)\tau(p,e) v_p(x),
\end{equation} where $\pi(p,e)$ is a polynomial in the operator $D$
which denotes the encoding, i. e.  they express the additional
delays introduced for each flow path at the node $start(e)$. The
term $\tau(p,e)$ denotes a function of the operator $D$, which we
will call the transfer function from $p$ to $e$ and which accounts
for the natural delay that, as previously noted, is inherent to the
transmission on each edge. The way to compute this function will be
explained in detail for each case. The process of determining the
network code consists of finding a set $\{\pi(p,e), e\in E, p\in
P(e)\}$.

{\bf Remark:} To simplify the encoding, we propose to use only
monomial $\pi(p,e)$'s, that is, $\pi(p,e) = D^{i_e(p)}$ for some
$i_e(p) \in \mathbb{Z}_0^+$. Thus the encoding consists of simply
adding the encoding vectors of all predecessor edges, each one
artificially delayed for zero or more time units as necessary. The
algorithm can be straightforwardly modified to allow all polynomials
in ${\mathbb F}_2(D)$ to be coefficients of the encoding
combination. We have chosen the proposed scheme because it is
simpler and because in this way we stick to the flow path graph that
has been computed beforehand. Since there exists an encoding for any
flow path graph, once a flow path graph has been computed we want to
follow it and profit from that in order to reduce the computational
complexity of finding the encoding equations for each edge. With
this simplification, the encoding takes the form

\begin{equation} \label{eq_localencoding}
v_e(x)=\sum_{p\in P(e)}D^{i_e(p)}\tau(p,e) v_p(x), \end{equation}
where $i_e(p)\in \mathbb{Z}_0^+$ are the exponents that express the
additional delays introduced at the node $start(e)$.

The exponents  $i_e(p)$ do no depend on the time variable $x$. We
remark that employing a time invariant encoding, as in this case,
 ensures that streams of symbols from each source can be
pipelined through the network.


The \emph{global encoding equation} expresses the symbol $v_e(x)$ to
be transmitted on edge $e$ in terms of the source symbols:
\begin{equation} \label{eq_globalencoding}
v_e(x)=\sum_{i=1}^h\sum_{y\in{\mathbb
Z}_0^+}\alpha(e,i,y)\sigma_i(x-y),
\end{equation} where $\alpha(e,i,y)\in {\mathbb F}_2 $ is the coefficient that specifies
the influence of each source symbol on $v_e(x)$.

Since $\sigma_i(z)=0$ for all $z<0$, the sum over $y$ in
(\ref{eq_globalencoding}) is finite, since the terms with $y>x$ will
all be 0.

Another way of expressing the global encoding equation, more useful
for our purposes, is by means of the field of rational functions
${\mathbb F}_2(D)$, where $D$ is the delay operator.

\begin{equation} \label{eq_globalencoding2}
v_e(x)=\sum_{i=1}^h F_{e,i}(D)\sigma_i(x),
\end{equation}
where $F_{e,i}(D)\in {\mathbb F}_2(D)$. Again, all the monomials in
$D^y$ with $y>x$ will give null terms in the sum.

The goal of the LIFE* algorithm is to determine the local encoding
equations of each edge $e$ in each flow path, and the global
encoding equation of the last edge in each flow path. In fact,
during the course of the algorithm, all the global encoding
equations will be determined.

In particular, the global encoding should be such that each sink $t$
can, after a certain constant delay $d_t$, extract all the $h$
information symbols belonging to a generation from the $h$ sources.
Each sink $t$ receives $h$ symbols, namely $\{v_{e_i}(x) \; | e_i\in
f^{s_i,t}, t=end(e_i), i=1,\ldots, h \}$. If the corresponding $h$
global encoding equations are linearly independent over the field
$\mathbb{F}_2(D)$, the sink $t$ can reconstruct the $h$ input
symbols of generation $x$ at time $x+d_t$, by solving (in a simple
way) the corresponding set of equations.

In the case of a flow acyclic network, all the edges in $E$ can be
visited according to the partial order $\prec$ induced by the flow
paths, starting by the edges with no predecessor ($ e\in E \; |\;
start(e)=s_i$ for some $i$), and proceeding in such a way that an
edge $e$ will not be visited until all its predecessors (all $p\in
P(e)$) have already been visited and their global encoding
equations computed. If the global encoding equations of all
predecessors of $e$ are substituted into (\ref{eq_localencoding}),
we get the global encoding equation for $e$.

When the network is flow cyclic the algorithm comes to a point at
which no more edges can be visited in topological order, that is,
the algorithm hits the flow cycle. Once the edges taking part in
that cycle have been identified, the encoding of the whole cycle
has to be treated as a whole. Here the old symbol removal
principle is useful. An appropriate version of Mason's formula
(see \cite{LinCostello}) for the transfer function on a circuit
will be the main tool to deal with it. We will explain this in
detail in the second part of this section.

Similar to the LIF and the LIFE algorithms, the new LIFE* algorithm
proceeds by maintaining, through the iterations of the main loop,
the following sets for each $t\in T$:

\begin{itemize}
  \item A set $E_t\subset E$, $|E_t|=h$, such that $E_t$ contains \emph{the most recently visited edge} on
each flow path in $f^t$.
  \item An $h\times h$  matrix $M_t$,
  in which the element $i,j$ will be the coefficient of
  $\sigma_i(x)$ in the global encoding equation of the $j$-th
  element  of the subset $E_t$.
  \[M_t=\left( \begin{array}{ccc}
  F_{e_{t,1},1}(D) & \cdots & F_{e_{t,h},1}(D)\\
  \vdots & & \vdots \\
F_{e_{t,1},h}(D) & \cdots & F_{e_{t,h},h}(D)
\end{array}\right)\]
where $e_{t,1},\ldots, e_{t,h}$ are the edges in $E_t$.

  \end{itemize}

Through each step of the algorithm we will impose the \emph{{\bf
full rank condition:}} \emph {The matrix $M_t$ must have rank $h$,
for all $t\in T$.} This condition is sufficient, but not
necessary, for obtaining a valid network code. At the final step,
for each $t\in T$ the set $E_t$ will be that of the $h$ edges
which arrive in $t$. If these edges carry the symbols $r_{t,1}(x),
\ldots, r_{t,h}(x)$ at time $x$, we have
\[\left(\sigma_1(x),\ldots,\sigma_h(x)\right)M_t=\left(r_{t,1}(x),\ldots,r_{t,h}(x)\right).\]
By the full rank condition the matrix $M_t$ is invertible,
and the symbols $\left(\sigma_1(x),\ldots,\sigma_h(x)\right)$ can be
recovered from the received ones as
\[\left(\sigma_1(x),\ldots,\sigma_h(x)\right)=\left(r_{t,1}(x),\ldots,r_{t,h}(x)\right)M_t^{-1}.\]

We will now explain how the algorithm will proceed.

\subsection{The flow acyclic parts of the network}\label{acyclic
part}

Whenever there is an edge $e$ that can be visited in topological
order (that is to say, one for which all the predecessors have been
already visited), the algorithm will proceed to update the sets
$E_t$ and $V_t$ in the following manner:

\[E_t:=\left\{ E_t \setminus
\left\{ f_{\leftarrow}^t(e) \right\} \right\} \cup \{e\}, \; {\rm
for \; each}\; t \in T(e)\]

We consider the general local encoding equation for edge $e$

\[v_e(x)=\sum_{p\in P(e)}D^{i_e(p)}\tau(p,e) v_p(x)\]

where $i_e(p)\in \mathbb{Z}_0^+$ are the unknowns and represent the
extra delay added at edge $e$ to maintain the full rank invariant.

In the acyclic case with unit link delay, $\tau(p,e)=D, \; \forall
p\in P(e)$. If the natural link delay is any other thing, even not
necessarily the same for all links, it will be straightforward to
adapt the corresponding equations. The local encoding formula
(\ref{eq_localencoding}) in the acyclic case with unit link delay
takes the form

\begin{equation}
\label{eq_localencoding2}
 v_e(x)=\sum_{p\in P(e)}D^{i_e(p)}D
v_p(x)=\sum_{p\in P(e)}D^{i_e(p)+1} v_p(x) \end{equation}

When the global encoding equations for all the $v_p(x)$, already
determined in the previous steps of the algorithm,  are substituted
in the above expression, we will have the global encoding equation
for $e$.

Further, for each $t \in T(e)$, replace in each matrix $M_t$ the
column corresponding to the encoding equation of
$f_{\leftarrow}^t(e)$ with the new one $v_e(x)$.

The unknowns will be chosen to have values that satisfy the full
rank invariant for all the matrices $M_t, t\in T(e)$, that have been
updated following the update of the corresponding set $E_t$.

\begin{conjecture}
There exists a finite value $I$, that depends on the graph and in
particular on the set $T(e)$, so that some set  $\{i_e(p): p\in P(e)
\mbox{ and } i_e(p)< I\}$, when applied to (\ref{eq_localencoding})
and (\ref{eq_localencoding2}), will satisfy the full rank condition.
\end{conjecture}

This conjecture is in a way similar to Lemma 6 in \cite{SET}. There
the field size required to guarantee the full rank invariant is
proven to be $|T(e)|$,
that is to say, each coefficient of the linear combination can be
chosen among $|T(e)|$ possibilities. If we assume $I = |T(e)|$, the
coefficients of the combinations in (\ref{eq_localencoding}) can be
chosen also among $|T(e)|$ different possibilities, namely $D^0=1,
D, D^2, \ldots, D^{|T(e)|-1}$. Nevertheless, the proof cannot use
the Linear Algebra arguments used there because the set with which
we are working is not a vector space.


The conjecture is further supported by Theorems 1 and 3 in
\cite{FrancisITW07}, and by software simulations for random networks
that we have carried out. We omit the details of these simulations.

For networks with random structure most encodings need no extra
delay at all, and in the few remaining cases a delay of one unit in
one of the incoming paths is enough to solve the problem for most of
them. This observation also tells us that in order to find the
encoding for each edge, which means finding the delay exponents
$i_e(p)$, an efficient approach will be to use a greedy algorithm
(see \cite{jaggi}) that starts by considering $i_e(p)=0, \; \forall
p\in P(e)$, checks if the full rank condition is satisfied, and in
case it is not, proceeds to increment the delay exponents one by one
until a solution is found. According to our simulations, the average
number of tries needed to find the solution for each edge will be
very low.

\subsection{Dealing with flow cycles}\label{cyclic part}

When the algorithm encounters a flow cycle the set of edges forming
part of the cycle has to be computed. Let us call $C_E$ the set of
edges and $C_V$ the set of nodes that are ends of those edges in the
cycle.

Let us call $P(C)$ the set of predecessor edges of the cycle,

 \[ P(C)=\{e \in E \setminus C_E\;|\;e\in P(e') \; \mbox{for some} \; e'\in C_E\}.\]

 We assume that the encoding equations for all the edges
$e \in P(C)$ have been already determined in previous steps of the
algorithm. The goal of this step of the algorithm will be finding at
once the encoding equation of all the edges in $C_E$. In a sense it
is as if the cycle as a whole is being treated as a kind of
'superedge'. All the individual edges in it will have basically the
same structure of equation, which will be a combination of the
equations of the edges in $P(C)$, that is

\begin{equation}
\label{eq_localcycle}
 v_C(x)=\sum_{p\in P(C)}D^{i_C(p)}\tau(p,C)
v_p(x)
\end{equation}

Here $\tau(p,C)$ is a notation with which we simply mean a transfer
function that will have to be computed separately for each
particular edge in $C_E$. 
Thus, each edge in $C_E$ will have a slightly different version of
that basic structure due to the fact that they lay in different
parts of the cycle and will consequently observe the incoming
equations with different delay.
 We will
distinguish two cases, namely, when the flow cycle is simple, or
when it is a knot. The distinction will be made just for the clarity
of explanation, since the simple case is just a particular case of
the knot case.

Once the exact encoding equation has been determined for all the
edges in $C_E$, the full rank invariant has to be checked only for
the last edges in the cycle for each flow path, that is to say, for
each $t$ with $f^t\cap C_E\neq \emptyset$ the corresponding full
rank condition must be satisfied by the  edge $e\in f^t\cap C_E$
such that $f_{\rightarrow}^{t}(e)\not\in C_E$.

\subsubsection{The simple flow cycle case}\label{simple cycle}

Suppose $C_E=\{e_1,e_2,\ldots,e_k\}$ with $end(e_i)=start(e_{i-1})$
for $i=1,\ldots,k-1$ and $end(e_k)=start(e_1)$.

Let us consider the local encoding equation of the cycle
(\ref{eq_localcycle}). We show now how to use the \emph{Old symbol
removal} principle.

We will focus on a certain edge in $C_E$, for instance $e_1$.
Suppose the local (and global) encoding equations of $e_k$, the
predecessor of $e_1$ in the cycle, have been determined exactly.

\[v_{e_k}(x)=\sum_{p\in P(C)}D^{i_{e_k}(p)}\tau(p,e_k) v_p(x)\]

If $e_1$ had no predecessor outside the cycle, that is to say
$P(e_1)=\{e_k\}$, then the encoding equation of $e_1$ would simply
be $v_{e_1}(x)=Dv_{e_k}(x)=\sum_{p\in
P(C)}D^{i_{e_k}(p)}\tau(p,e_k)D v_p(x)$, which means
$\tau(p,e_1)=D\tau(p,e_k) \; \forall p\in P(C)$. In the same way it
is clear in general that if $p\in P(C)\setminus P(e_1)$, then
$\tau(p,e_1)=D\tau(p,e_k)$.

Now suppose that there is one particular edge $p_1\in P(C)$ which is
the unique predecessor of $e_1$ not in $C_E$, this implies
$\tau(p_1,e_1)=D$. Not removing at that point the old contribution
of that predecessor would mean that the new contribution would mix
with the old ones on each loop of the cycle and would keep
circulating forever. In order to avoid this we want to remove the
old contribution that came on edge $p_1$ $k$ time instants ago
(where $k$ is obviously the length of the cycle) and contribute to
the circulation in the cycle with only the newest symbol coming on
$p_1$. This is done as follows

\[\begin{array}{ll}
v_{e_1}(x) & =D\cdot
v_{e_k}(x)+D^{i_C(p_1)}\tau(p_1,e_1)\left[-v_{p_1}(x-k)+
v_{p_1}(x)\right] \\
 & =D\cdot
v_{e_k}(x)+D^{i_C(p_1)}D\left[-v_{p_1}(x-k)+ v_{p_1}(x)\right]
\end{array}\]

Here we have used the minus operator $(-)$, despite all the
operations are always on the binary field, in order to emphasize
which symbols are being removed from the circulation.

In general, if $e\in C_E$ has several predecessors not lying in the
cycle, the local encoding equation of $e$ in terms of the
predecessors of $e$ takes the form

\[\begin{array}{ll}
v_{e}(x) & =D\cdot v_{p_C(e)}(x)+ \sum_{p\in P(e)\cap
P(C)}{D^{i_C(p)}\tau(p,e)\left[-v_{p}(x-k)+ v_{p}(x)\right]}\\
 & =D\cdot v_{p_C(e)}(x)+ \sum_{p\in P(e)\cap
P(C)}{D^{i_C(p)}D\left[-v_{p}(x-k)+ v_{p}(x)\right]}\end{array}\]
where $p_C(e)=P(e)\cap C_E$.

The result of doing this at the entrance in the cycle of each
predecessor $p\in P(C)$ is that only one `instance' of the symbols
carried by each $p\in P(C)$ will be circulating on each edge $e\in
C_E$. It is easy to see that the transfer function from each
predecessor of the cycle to each edge in the cycle will be
$\tau(p,e)=d(p,e) \; \forall p\in P(C), e\in C_E$, where $d(p,e)$ is
the `distance' measured in number of edges in the cycle that lay
between $end(p)$ and $end(e)$.

To summarize, the local encoding equation of each edge $e\in C_E$ in
terms of the predecessors of the cycle is

\[v_e(x)=\sum_{p\in P(C)}D^{i_C(p)+d(p,e)} v_p(x)\]
where again, the only unknowns are the values $i_C(p)\in
\mathbb{Z}_0^+$.

An example of the use of this procedure can be found in the Appendix
when encoding the network of Example 3.

\subsubsection{The knot case}\label{knot}

Suppose the set $C_E$ is not just a simple flow cycle but forms a
knot.

Again the local encoding equation for the whole knot will share a
common structure
\[v_C(x)=\sum_{p\in P(C)}D^{i_C(p)}\tau(p,C) v_p(x)\]

Once more the idea is to treat the whole knot as a kind of
'superedge'.

Here the main idea is the same as before: the old symbols must be
removed from the circulation. In order to do it one needs to know
how those arrive at each edge of the knot, and for this we need as a
tool Mason's formula (see
\cite{LinCostello},\cite{MasonPaperMengetal}) for the computation of
the transfer function on a cyclic circuit.

We apply Mason's formula to the  directed line graph associated with
$C_E$ in the following way: Two edges $e'$ and $e$ in $C_E$ are
considered adjacent (and an arc starting in $e'$ and ending in $e$
will be drawn in the line graph) if and only if there exists a flow
$f^t$ for some $t\in T$ such that $f_{\leftarrow}^{t}(e)=e'$.

For each symbol entering the knot we have to compute the
corresponding transfer function over all the edges in $C_E$. For
this we will consider the entrance point of the symbol (the edge at
which that symbol enters) and the exit point (that is, the edge at
which we want the transfer function of that symbol), and will apply
Mason's formula  between these points in the line graph above
mentioned and using the delay operator $D$ as the branch gain (see
\cite{LinCostello}) of each edge. Again, each branch gain can be
taken to be whatever function of $D$ models best the actual behavior
of the transmission on that edge and the corresponding equations can
be adapted consequently.

 We show  in detail how to compute the transfer functions by means of
 Example~4 in the Appendix.

The local encoding equation of edge $e\in C_E$ in terms of its
predecessors is

\begin{equation}
v_e(x)=\sum_{p\in P(e)\cap C_E}D v_p(x)+D\left[\sum_{p\in P(e)\cap
P(C)} {D^{i_C(p)}v_p(x)}-\sum_{e'\in P(e)\cap C_E}\sum_{p'\in
P'(e)\cap P(C)}{ D^{i_C(p')}\tau(p',e')v_{p'}(x)}\right]
\end{equation}

where $P'(e)=\{p\in E\; | \; end(p)=start(e)\}$. Note that
$P(e)\subseteq P'(e)$ but the converse is not true in general. For
instance, in Example 4 $P'(e_{13})=\{e_5,e_8,e_{17}\}$ while
$P(e_{13})=\{e_5,e_{17}\}$ (see Figure~6).

The second sum in the formula brings the updated versions of the
symbols that enter the knot at that point, while the double sum in
the third term of the formula takes care of removing the old
symbols.

 This results in the
following local encoding equation of each edge in the cycle in terms
of the predecessors of the cycle:

\begin{equation}
v_e(x)=\sum_{p\in P(C)}D^{i_C(p)}\tau(p,e)v_p(x)
\end{equation}

One can see that the previous case is just a particular case of this
one, since the line graph that will be associated to a simple flow
cycle will always contain a simple cycle itself, and the
corresponding transfer function between each $p\in P(C)$ and each
edge $e$ in the cycle will be $\tau(p,e)=d(p,e)$ as was shown in
\ref{simple cycle}.

A final observation at this point is that when the network is flow
acyclic or contains only simple flow cycles, the global encoding
equations will only contain polynomials on $D$, and not rational
functions. Rational functions will be the result of using Mason's
formula on knots.

Considerations about how to decode will be discussed in the
Appendix.

We conclude the current Section by summarizing the complete LIFE*
algorithm.

\noindent {\bf ALGORITHM LIFE* }
\begin{description}
\item[Input:]

A directed multigraph $G$; a set of flow paths $f$.
\item[Initializing:]

$\forall t:$
\begin{itemize}
    \item $E_t = \{e | e \in f^{s_i,t}, start(e)=s_i, i=1,\ldots,
h\},$
    \item $\{v_e(x) = D\sigma_i(x)=\sigma_i(x-1) | \; e\in
 E_t, start(e) = s_i \}$ or, equivalently,  $M_t =DI_h,$ $ \forall t\in
 T$.
\end{itemize}



\item[Main loop:]

Select an edge $e$ for which the encoding equations have not yet
been determined, but for which the global encoding equations of all
the predecessor edges in $P(e)$ have been determined.  Then proceed
with the update of the set of current edges and current encoding
equations as described in Subsection~\ref{acyclic part}.

If selecting such and edge is not possible, then a flow cycle has
been encountered. Follow the procedure explained in
Subsection~\ref{cyclic part}.

\item[Output:]
For each edge, the local encoding as given by
(\ref{eq_localencoding}) is produced. At the end of the algorithm,
$E_t=\{e \; | \; end(e)=t\}$ and $V_t$ is still a set of $h$
linearly independent equations from which $t$ can recover the input.

\end{description}

Examples of application of the algorithm can be found in the
Appendix.

\section{Practical considerations}\label{sec_considerations}

The algorithm will execute the main loop at most $|E|$ times. The
exact complexity depends on details of the algorithm not discussed
here. However the complexity of the LIF and LIFE algorithms are
similar. For discussions on the complexity of the algorithm we also
refer the reader to \cite{SET}.

The encoding presented here follows a flow path graph given for a
network. This flow path graph is not necessarily unique and the
choice made when computing the flow path graph determines much of
the possible encodings that can be achieved. Which flow path graph
is the best choice remains an open problem. First one should
consider in which way the solution wants to be optimal (minimal
delay, minimal number of link used, minimal number of encoding nodes
...). Some notions of minimality in the flow path graph can be
considered that we will not discuss here. Also we will not discuss
the different strategies that can be used in order to compute a flow
path graph.

Once the flow path graph for the network has been computed, the
algorithm proceeds by following a topological order of the edges
whenever that is possible (until a flow cycle or knot is found).
However, this topological order is not unique. In many cases there
is a certain choice to be made at each step about which edge will be
encoded next of the several that follow in the order. This choice
might in certain cases influence the total amount of delay necessary
for the encoding. Examples can be given in which different choices
of order lead to different final amounts of delay. Which ordering is
most convenient for each  flow path graph is also an open problem.

Another consideration to take into account is that the presence of
added delay means that the nodes at which the delay has to be
introduced must have memory elements to store the symbols that have
to be 'artificially' delayed. In most cases the sinks will need to
use memory in order to be able to solve the equations. In any case
the maximum delay used is finite. In case no extra delay needs to be
added to the maximum delay needed on each path will correspond to
the total length of that path from source to sink.


The OSR and PI principles are also not necessary, but they help to
keep the encoding simpler. Still, encodings can be found for flow
cyclic networks in which the principles are not respected. Not
respecting the PI principle is in fact equivalent to choosing a
different flow graph path.

\subsection{Network precoding}
The inverse matrix of the encoding equation system may contain
rational functions with denominators not on the form of $D^i$, for
some constant $i$. If so, the encoding is 'catastrophic' in the
sense that an error occurring in one of the transmissions can result
in an infinite sequence of errors at the decoding sink. In order to
avoid that, once the encoding has been computed using the LIFE*
algorithm, one can compute the polynomial which is maximum common
divisor of all the rational functions resulting in the encoding
process and introduce a pre-coding of the symbols generated by the
sources, multiplying them by that maximum common divisor before they
are introduced in the network. Alternatively, we can carry out this
precoding locally in the nodes where a path enters a knot. We omit
the details.

After this precoding is introduced, the network code as viewed from
the perspective of the sink is polynomial, and any error that might
occur will cause only a limited error propagation that can be
handled by a suitable error correcting or erasure restoring code.

\section{Conclusions}\label{sec_conclusions}

The LIFE* algorithm is able to encode any given network with
polynomial complexity and over the binary field. The addition of
delay at some nodes is not a major drawback. In fact, any network
encoding will in practice have intrinsic delay associated with it,
and the delay will differ over the various paths.
Thus in most cases, LIFE* does not need to introduce extra delay.
In the few cases in which we actually need to introduce extra delay,
this extra delay is what allows us to get the encoding on the binary
field, which would have been impossible otherwise.
For networks where the LIF/LIFE algorithms work, LIFE* will perform
with essentially the same complexity as the others, i. e., there is
no known more efficient algorithm in these cases. If the network
contains knots, which many practical networks do, no other known
algorithm works, but for LIFE* the complexity may become dominated
by the calculation of Mason's formula. The greedy approach to
finding the coding coefficients for each edge performs essentially
as in the acyclic cases, also for knots.

\appendix
\section*{Appendix: Examples}

We will show here how the LIFE* algorithm will find encodings for
the different types of networks shown in Figure~2.

\subsection*{Example 1}

The $(2,4)$ combination network is presented in Figure~2 a).  It is
known (\cite{SET},\cite{soljanin}, \cite{cyclopathic}) to be a
network
 which when extra delay is not used requires a
finite field larger than $\mathbb{F}_2$. We show here how the LIFE*
algorithm will work on the flow graph given in
Figure~3.

\begin{figure}[htbp]
\centerline{\includegraphics[width=9cm,keepaspectratio=true,angle=0]{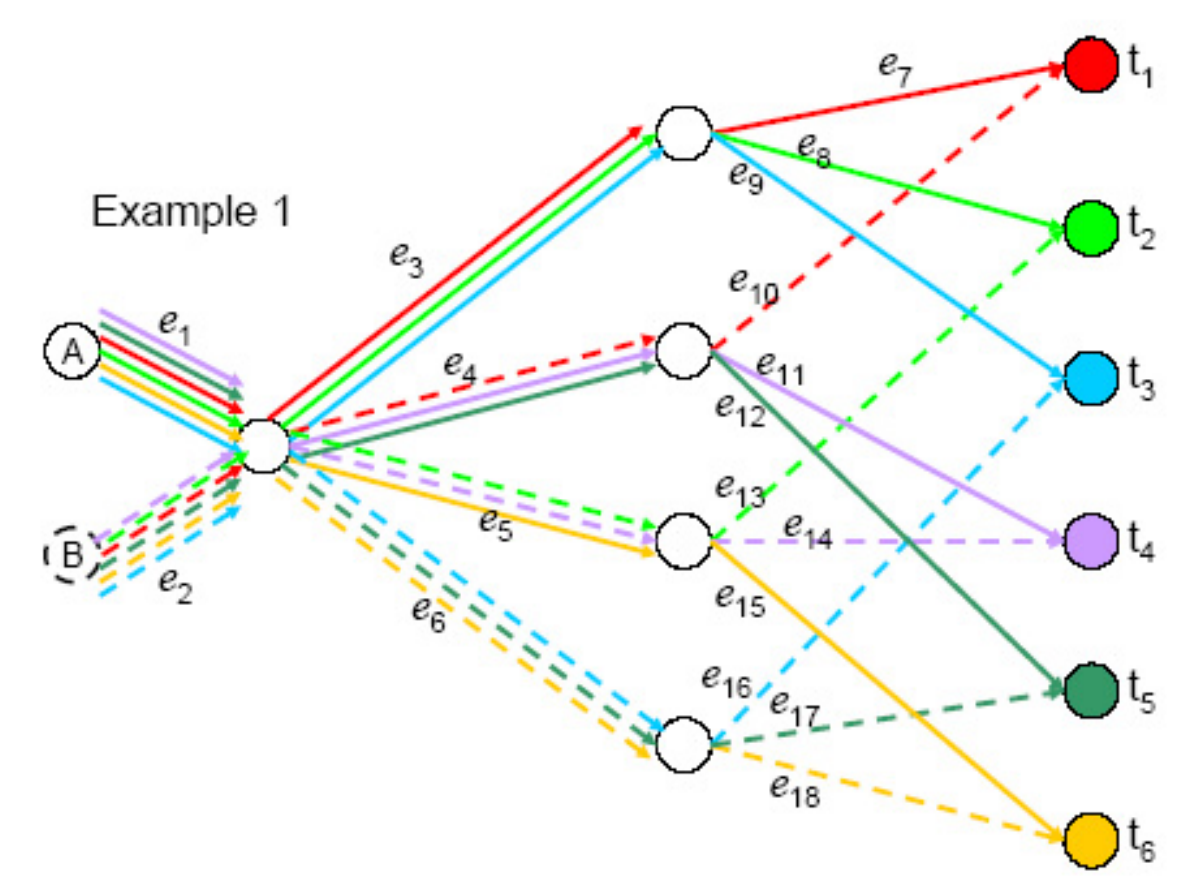}}
\label{Fig_Example1} \caption{A flow path graph for the $(2,4)$
combination network.}
\end{figure}

In order to better follow the progress of the algorithm, we have
assigned labels $e_1, \ldots, e_{18}$ to the edges in the network
following a topological order. Also, for simplicity in the notation
we have called the two sources $A$ and $B$, and $a(x)$ and $b(x)$
are the binary symbols released by the sources at time $x$.

The flow paths are represented with a code of colors and patterns in
order to make is visually easy to follow. Each sink has a color
assigned and each of the 2 sources has a pattern assigned (solid for
$A$, dashed for $B$). The flow path from a source to a sink will be
drawn in the color of the sink and with the pattern of the source.

The LIFE* algorithm starts by setting the following initial values:

\[\begin{array}{l}
E_{t_1}=E_{t_2}=\cdots =E_{t_6}=\{e_1,e_2\}\\
v_{e_1}(x)=Da(x)=a(x-1)\\
v_{e_2}(x)=Db(x)=b(x-1)
\end{array}\]

Hence
\[M_{t_i}=\left(\begin{array}{cc}
D & 0\\ 0 & D \end{array}\right), \; i=1,\ldots, 6.\]

Now the algorithm enters the main loop:

\begin{itemize}

\item For encoding $e_3$ we can observe that the only predecessor
is $e_1$, and thus

 \[v_{e_3}(x)=D^{i_{e_3}(e_1)+1}v_{e_1}(x)=D^{i_{e_3}(e_1)+2}a(x)\]

 Edge $e_3$ is in the flows to sinks $t_1, t_2$ and $t_3$, so we
 update the corresponding sets of edges and  matrices

\[E_{t_1}=E_{t_2}=E_{t_3}=\{e_3,e_2\}\]

\[M_{t_1}=M_{t_2}=M_{t_3}=\left(\begin{array}{cc}
D^{i_{e_3}(e_1)+2} & 0\\ 0 & D \end{array}\right)\]

Clearly the choice $i_{e_3}(e_1)=0$ makes all the matrices non
singular. Thus the encoding of $e_3$ is
\[v_{e_3}(x)=Dv_{e_1}(x)=D^2a(x)=a(x-2)\]

\item $e_4$ has two predecessors, $e_1$ and $e_2$.

 \[v_{e_4}(x)=D^{i_{e_4}(e_1)+1}v_{e_1}(x)+D^{i_{e_4}(e_2)+1}v_{e_2}(x)=D^{i_{e_4}(e_1)+2}a(x)+D^{i_{e_4}(e_2)+2}b(x)\]

Edge $e_4$ is in the flows to sinks $t_1, t_4$ and $t_5$, so we
 update the corresponding sets of edges and  matrices

\[E_{t_1}=\{e_3,e_4\},E_{t_4}=\{e_4,e_2\},E_{t_5}=\{e_4,e_2\}\]

\[M_{t_1}=\left(\begin{array}{cc}
D^2& D^{i_{e_4}(e_1)+2}\\ 0 & D^{i_{e_4}(e_2)+2}
\end{array}\right),
M_{t_4}=M_{t_5}=\left(\begin{array}{cc}
 D^{i_{e_4}(e_1)+2} & 0 \\
D^{i_{e_4}(e_2)+2} & D
\end{array}\right),\]

Again one can see that the choice $i_{e_4}(e_1)=i_{e_4}(e_2)=0$
makes all three matrices non singular. Hence

\[v_{e_4}(x)=Dv_{e_1}(x)+Dv_{e_2}(x)=D^2a(x)+D^2b(x)=a(x-2)+b(x-2)\]

\item Edge $e_5$ has $e_1$ and $e_2$ as predecessors, and the
form of the encoding is

\[v_{e_5}(x)=D^{i_{e_5}(e_1)+1}v_{e_1}(x)+D^{i_{e_5}(e_2)+1}v_{e_2}(x)=D^{i_{e_5}(e_1)+2}a(x)+D^{i_{e_5}(e_2)+2}b(x)\]

Edge $e_5$ takes part in the flows to sinks $t_2, t_4$ and $t_6$,
and the corresponding updating of edge sets and matrices is as
follows:

\[E_{t_2}=\{e_3,e_5\}, E_{t_4}=\{e_4,e_5\}, E_{t_6}=\{e_5,e_2\}\]

\[M_{t_2}=\left(\begin{array}{cc}
D^2& D^{i_{e_5}(e_1)+2}\\ 0 & D^{i_{e_5}(e_2)+2}
\end{array}\right),
M_{t_4}\left(\begin{array}{cc}
 D^2 & D^{i_{e_5}(e_1)+2} \\
D^2 & D^{i_{e_5}(e_2)+2}
\end{array}\right),
M_{t_6}\left(\begin{array}{cc}
  D^{i_{e_5}(e_1)+2} & 0 \\
 D^{i_{e_5}(e_2)+2} & D
\end{array}\right),\]

Now clearly any value of $i_{e_5}(e_1)$ and $i_{e_5}(e_2)$ will make
matrices $M_{t_2}$ and $M_{t_6}$ non singular, but in order to get
$M_{t_4}$ non singular we need those two values to be different,
hence setting both equal to 0 does not work in this case. A possible
choice would be $i_{e_5}(e_1)=1$, $i_{e_5}(e_2)=0$, which gives us
the next encoding.

\[v_{e_5}(x)=D^2v_{e_1}(x)+Dv_{e_2}(x)=D^3a(x)+D^2b(x)=a(x-3)+b(x-2)\]

\item In the same manner we work with edge $e_6$, which has only one
predecessor, namely $e_2$ and following the same procedure as before
we can see that setting the only unknown exponent to 0 will give a
correct encoding.

\[v_{e_6}(x)=Dv_{e_2}(x)=D^2b(x)=b(x-2)\]

\end{itemize}

{\bf Remark:}

The particular case we have seen in the encodings of edges $e_3$ and
$e_6$, that is to say, an edge with only one predecessor is always
solved in the same manner, copying the symbol carried by the
predecessor and adding the natural delay unit. This corresponds to

\[v_e(x)=Dv_p(x)\]
when $p$ is the only predecessor of $e$, that is to say,$P(e)=\{p\}$
. (This means that the exponent $i_e(p)$ has been chosen to be 0.)

The updated matrices will keep full rank since, for each $t\in T(e)$
the corresponding updated matrix will be the result of multiplying
by $D$ the elements of one of the columns of the old matrix, which
does not alter the rank of the matrix. $\square$

The rest of the encoding steps in this example are trivial in that
sense, since all the rest of the edges have an only predecessor.

To complete the example we will illustrate how sinks can decode,
this will also show what the delay means at the receiver end.

Let us focus on sink $t_6$. According to the encoding just computed
this sink will receive at time $x$ the symbols $a(x-4)+b(x-3)$ and
$b(x-3)$.

Since we are assuming $a(x)=b(x)=0$ for all negative $x$, sink $t_6$
will receive zeros on both channels until time $x=3$ in which it
receives $0+b(0)$ and $b(0)$. This obviously gives him the knowledge
only of symbol $b(0)$. But at time $x=4$ it receives $a(0)+b(1)$ and
$b(1)$. The knowledge of $b(1)$ allows it to recover $a(0)$, which
completes the recovery of the symbols of generation 0. Proceeding in
the same way it will complete the recovering of the symbols of
generation $x$ at time $x+4$. The total delay observed by sink $t_6$
is 4, which in this case coincides with the maximum power of $D$
used in the encoding equations arriving in $t_6$.

This can also be interpreted in terms of matrices.

\[M_{t_6}\left(\begin{array}{cc}
  D^4 & 0 \\
 D^3 & D^3
\end{array}\right),\]
 If we create a vector with the symbols that arrive at $t_6$ at time $x$
 and denote it as  $[r_{t_6,1}(x),r_{t_6,2}(x)]$,  the encoding
 process can be described as

 \[[a(x),b(x)]M_{t_6}=[r_{t_6,1}(x),r_{t_6,2}(x)]\]

 (which is equivalent to saying that
 $r_{t_6,1}(x)=a(x-4)+b(x-3),r_{t_6,2}(x)=b(x-3)$).

 Now the decoding process can be described as

 \[[a(x),b(x)]=[r_{t_6,1}(x),r_{t_6,2}(x)]M_{t_6}^{-1}=[r_{t_6,1}(x),r_{t_6,2}(x)]\frac{1}{D^4}\left(\begin{array}{cc}
 1 & 0 \\ 1 & D \end{array}\right)\]

 that is to say

\[\begin{array}{l}
a(x)=r_{t_6,1}(x+4)+r_{t_6,2}(x+4)\\
b(x)=r_{t_6,2}(x+3) \end{array}\]

which again shows how, to recover the symbols in generation $x$,
sink $t_6$ has to wait until receiving symbols at time $x+4$.

{\bf Remark:}

 In general the delay experienced by each sink is lower
bounded by the maximum length of the flow paths arriving at it from
the $h$ different sources and upper bounded by the maximum power of
the delay operator $D$ used in the global encoding equations of the
edges arriving at that sink. $\square$

The upper bound is not always tight. To illustrate this let us
consider the decoding that sink $t_4$ in the example has to do.
Despite the maximum power of $D$ for that sink is 4, it is easy to
see that $t_4$ will complete the recovery of generation $x$ at time
$x+3$. (But in addition it is absolutely necessary for $t_4$ to keep
one memory element in order to be able to decode.)

\subsection*{Example 2}

This example shows a network which is link cyclic but flow acyclic.
The encoding process will work analogous to what was shown in the
previous example.

\begin{figure}[htbp]
\centerline{\includegraphics[width=4cm,keepaspectratio=true,angle=0]{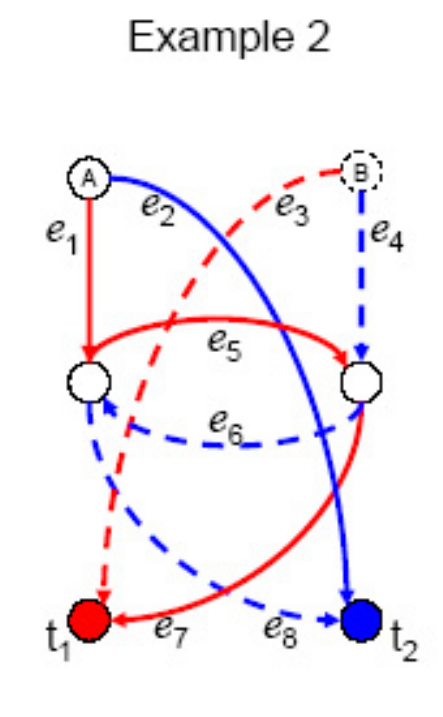}}
\label{Fig_Example2} \caption{The unique flow path graph for Example
2.}
\end{figure}

The edges in Figure~4 have been numbered according to a topological
order and one can observe that each edge has in fact only one
predecessor, hence the encoding becomes trivial. We simply show here
the result obtained.

\[\begin{array}{l}
v_{e_1}(x)=v_{e_2}(x)=a(x-1) \\
v_{e_3}(x)=v_{e_4}(x)=b(x-1) \\
v_{e_5}(x)=a(x-2) \\
v_{e_6}(x)=b(x-2) \\
v_{e_7}(x)=a(x-3) \\
v_{e_8}(x)=b(x-3)
\end{array}\]

As we can see, link cyclic but flow acyclic networks do not present
any additional problem for encoding, they behave exactly as the
acyclic networks did.

\subsection*{Example 3}

Here we deal with a flow cyclic network that contains a simple flow
cycle. Figure~5 shows the unique flow path graph for this network.

\begin{figure}[htbp]
\centerline{\includegraphics[width=12cm,keepaspectratio=true,angle=0]{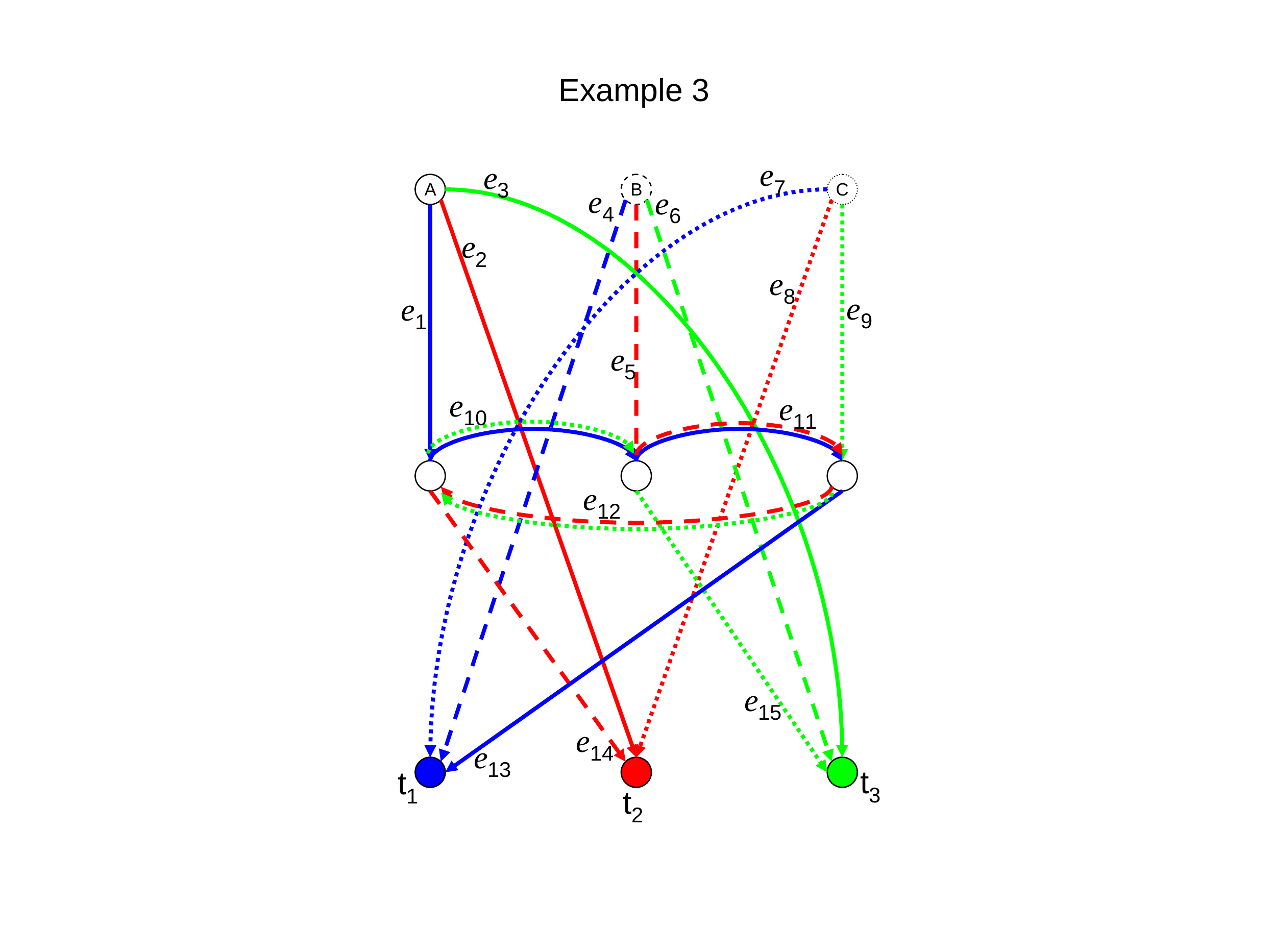}}
\label{Fig_Example3} \caption{The unique flow path graph for Example
3.}
\end{figure}

The initialization will give us the encoding of the first 9 edges

\[\begin{array}{l}
v_{e_1}(x)=v_{e_2}(x)=v_{e_3}(x)=a(x-1) \\
v_{e_4}(x)=v_{e_5}(x)=v_{e_6}(x)=b(x-1) \\
v_{e_7}(x)=v_{e_8}(x)=v_{e_9}(x)=c(x-1)
\end{array}\]

Now no more edges can be visited following a topological order. The
cycle has set of edges $C_E=\{e_{10},e_{11},e_{12}\}$ and set of
predecessors $P(C)=\{e_1,e_5,e_9\}$.

The structure of the local encoding equation in the cycle will be

\[v_C(x)= D^{i_C(e_1)}\tau(e_1,C)v_{e_1}(x)+D^{i_C(e_5)}\tau(e_5,C)v_{e_5}(x)+D^{i_C(e_9)}\tau(e_9,C)v_{e_9}(x)\]

Since we are in the case of a simple flow cycle, we can follow the
formula given in \ref{simple cycle} for the local encoding equation
of each of the three edges in $E_C$.

\[v_{e_j}(x)= D^{i_C(e_1)+d(e_1,e_j)}v_{e_1}(x)+
D^{i_C(e_5)+d(e_5,e_j)}v_{e_5}(x)+D^{i_C(e_9)+d(e_9,e_j)}v_{e_9}(x),\;
j=10, 11, 12\]

Inspection of the graph shows that

\[\begin{array}{l}
d(e_1,e_{10})=1, d(e_5,e_{10})=3, d(e_9,e_{10})=2\\
d(e_1,e_{11})=2, d(e_5,e_{11})=1, d(e_9,e_{11})=3\\
d(e_1,e_{12})=3, d(e_5,e_{12})=2, d(e_9,e_{12})=1 \end{array}\]

and the substitution in the above expression gives
\[\begin{array}{l}
v_{e_{10}}(x)= D^{i_C(e_1)+1}a(x-1)+D^{i_C(e_5)+3}b(x-1)+D^{i_C(e_9)+2}c(x-1)\\
v_{e_{11}}(x)= D^{i_C(e_1)+2}a(x-1)+D^{i_C(e_5)+1}b(x-1)+D^{i_C(e_9)+3}c(x-1)\\
v_{e_{12}}(x)=D^{i_C(e_1)+3}a(x-1)+D^{i_C(e_5)+2}b(x-1)+D^{i_C(e_9)+1}c(x-1)
\end{array}\]

The full rank invariant condition must be checked for edge $e_{10}$
in the flow to sink $t_3$, for edge $e_{11}$ in the flow to sink
$t_1$ and for edge $e_{12}$ in the flow to sink $t_2$. This gives us
the following matrices:

\[ M_{t_1}=\left(\begin{array}{ccc}
D^{i_C(e_1)+3} & 0  & 0\\
D^{i_C(e_5)+2}& D & 0\\
D^{i_C(e_9)+4} & 0 & D
\end{array}\right),
M_{t_2}=\left(\begin{array}{ccc}
D & D^{i_C(e_1)+4} & 0\\
0 & D^{i_C(e_5)+3}& 0\\
0 & D^{i_C(e_9)+2} & D
\end{array}\right),
M_{t_3}=\left(\begin{array}{ccc}
D & 0 & D^{i_C(e_1)+2} \\
0 & D & D^{i_C(e_5)+4}\\
0 & 0 & D^{i_C(e_9)+3}
\end{array}\right)\]

Clearly any value of $i_C(e_1),i_C(e_5)$ and $i_C(e_9)$ satisfies
the full rank invariant and we choose the simplest one setting the
three unknowns to be 0.

The global encoding equations of the edges in the cycle are as
follows

\[\begin{array}{l}
v_{e_{10}}(x)= a(x-2)+b(x-4)+c(x-3)\\
v_{e_{11}}(x)= a(x-3)+b(x-2)+c(x-4)\\
v_{e_{12}}(x)=a(x-4)+b(x-3)+c(x-2)
\end{array}\]

 The encoding now of edges $e_{13}$, $e_{14}$ and $e_{15}$ is
trivial since each of them has only one predecessor.

\[\begin{array}{l}
v_{e_{13}}(x)= a(x-4)+b(x-3)+c(x-5)\\
v_{e_{14}}(x)= a(x-5)+b(x-4)+c(x-3)\\
v_{e_{15}}(x)=a(x-3)+b(x-5)+c(x-4)
\end{array}\]

The delay at the final receivers is 4, even when the maximum
exponent of $D$ in the equations received by the sinks is 5. Besides
there is some extra memory needed in order to decode. For instance,
recovering the element $a(1)$ and hence completing the generation
$1$, can be done by sink $t_1$ at time $x=5$, provided it kept in
memory the element $c(0)$.

A slight modification could be done for the encoding of the edges
whose predecessors lie in the cycle, in such a way that they get the
last updated values, for instance, edge $e_{13}$ can benefit from
the fact that edge $e_9$ enters in the same node from which $e_{13}$
exits and hence get an updated version of the symbol carried by
$e_9$, then the encoding of $e_{13}$ would be

\[v_{e_{13}}(x)= a(x-4)+b(x-3)+c(x-2)\]
which is actually the same encoding that has the edge $e_{12}$, and
results in smaller memory needed at the receiver $t_1$.

{\bf Remark:}

  In general,using this last observation,  the local
encoding equation of an edge $e\not\in C_E$ with a predecessor
$p_C\in P'(e)\cap P(C)$ would be

\[v_e(x)=D^{i_e(s_C)}v_{s_C}(x)+\sum_{p\in P(e)\setminus
\{p_C\}}D^{i_e(p)}D v_p(x)\]

where $s_C$ is the successor of $p_C$ that lies in the cycle, that
is to say, the edge in $C_E$ with $start(s_C)=start(e)=end(p_C)$.

We finally remark that in a simple flow cycle, the element $s_C$ is
are unique, even in there are several elements $p_C$ in $P'(e)\cap
P(C)$. $\square$

\subsection*{Example 4}

In this example we show how to work with a knot. Figure~6 shows the
essentially unique flow path graph for the network given.

\begin{figure}[htbp]
\centerline{\includegraphics[width=15cm,keepaspectratio=true,angle=0]{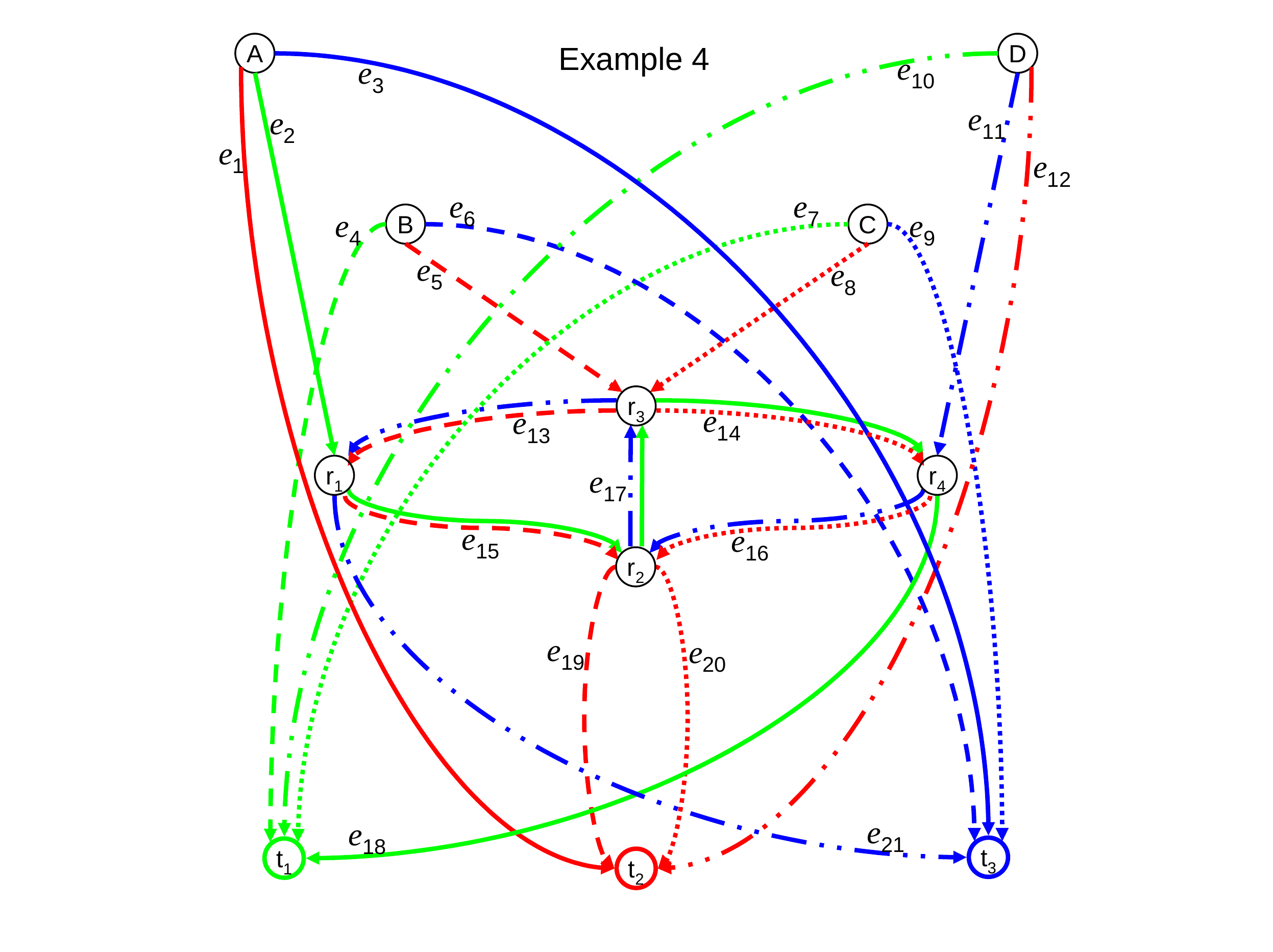}}
\label{Fig_Example4} \caption{A flow path graph for the network in
Example 4.}
\end{figure}

The initialization values are

\[\begin{array}{l}
v_{e_1}(x)=v_{e_2}(x)=v_{e_3}(x)=a(x-1)\\
v_{e_4}(x)=v_{e_5}(x)=v_{e_6}(x)=b(x-1)\\
v_{e_7}(x)=v_{e_8}(x)=v_{e_9}(x)=c(x-1)\\
v_{e_{10}}(x)=v_{e_{11}}(x)=v_{e_{12}}(x)=d(x-1)
\end{array}\]

No more edges can be visited in topological order because
$C_E=\{e_{13},e_{14},e_{15},e_{16},e_{17}\}$ form a flow cycle. In
fact it is a non simple cycle, since it contains two flow cycles
$e_{13}\prec e_{15} \prec e_{17} \prec e_{13}$ and $e_{14}\prec
e_{16} \prec e_{17} \prec e_{14}$, both sharing the edge $e_{17}$.
Hence we are in presence of a flow knot. (See Figure 7 a))

\begin{figure}[htbp]
\centerline{\includegraphics[width=12cm,keepaspectratio=true,angle=0]{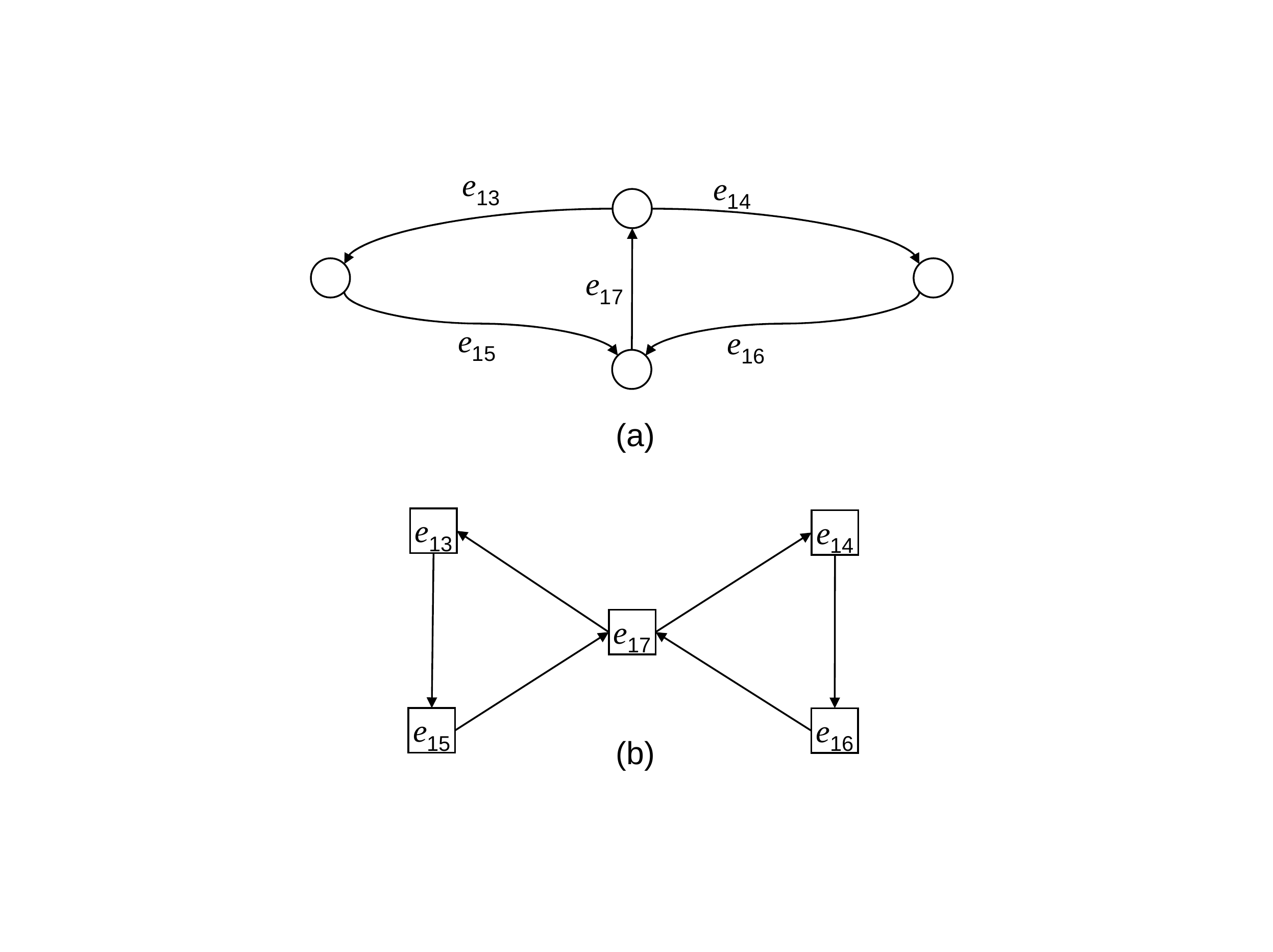}}
\label{Fig_Knot Example4} \caption{The knot in Example 4 and its
line graph.}
\end{figure}

The predecessors of the knot are $P_C=\{e_2,e_5, e_8, e_{11}\}$.

The general structure of the local encoding equation for the knot is

\[v_C(x)=D^{i_C(e_2)}\tau(e_2,C)v_{e_2}(x)+
D^{i_C(e_5)}\tau(e_5,C)v_{e_5}(x)+D^{i_C(e_8)}\tau(e_8,C)v_{e_8}(x)+
D^{i_C(e_{11})}\tau(e_{11},C)v_{e_{11}}(x)\]

In order to find the local encoding of each edge in the knot we have
to use Mason's formula on a line graph to compute the transfer
function for each of the symbols carried by the predecessors of the
knot.

The line graph is shown in  Figure 7 b). The nodes correspond to the
edges in $C_E$. The edge connecting $e_{13}$ to $e_{15}$ is drawn
because of the flow path from source $B$ to sink $t_2$, the edge
connecting $e_{15}$ to $e_{17}$ is determined by the flow from
source $A$ to sink $t_1$. In the same way we draw the other
connections in the line graph following the flow paths.

The branch gain of each connection is $D$. Mason's formula is as
follows
\[\tau(e_j,e_k)=\frac{\sum F_i(e_j,e_k) \Delta _i(e_j,e_k)}{\Delta}\]

where $\Delta=1+\sum c_i-\sum c_ic_j+ \cdots $, $F_i(e_j,e_k)$ is
the function corresponding to the $i-th$ forward path form $e_j$ to
$e_k$ and $\Delta_i(e_j,e_k)$ is defined as $\Delta$ but counting
only the cycles in the circuit that are disjoint with the $i$-th
forward path. Here we are using the notations in \cite{LinCostello},
and we refer the reader there for a more detailed explanation of
Mason's formula.

\begin{itemize}

\item
We will now focus on the symbol that enters through edge $e_2$ into
$e_{15}$. Its itinerary through the knot is shown in Figure 8 (a).

\begin{figure}[htbp]
\centerline{\includegraphics[width=12cm,keepaspectratio=true,angle=0]{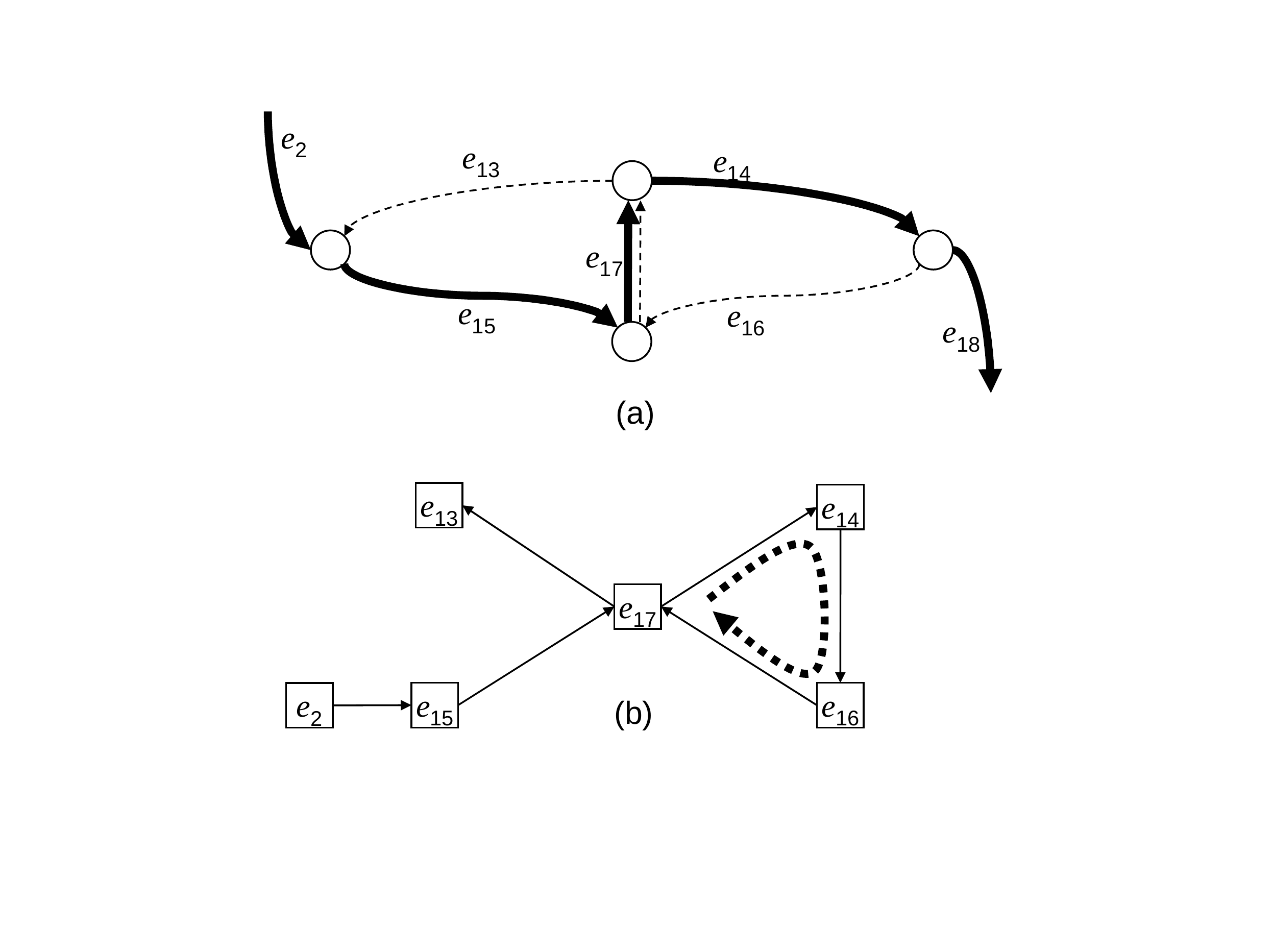}}
\label{Fig_Flow e2 Knot Example4} \caption{The itinerary of symbols
carried by $e_2$ in the knot of Example 4.}
\end{figure}

The edges used by the flow path that carries that symbol are
represented as bold lines, but we observe also that when the symbol
arrives at node $end(e_{17})$ this node will distribute it not only
to edge $e_{14}$, but  since there is  a flow path connecting edge
$e_{17}$ with edge $e_{13}$, the symbol in question will travel also
on edge $e_{13}$ and in the same way we can see that it will travel
also on edge $e_{16}$. That is represented in dashed lines in Figure
7 (a). On the other hand, using memory, the node
$end(e_2)=end(e_{13})=start(e_{15})$ can remove the contribution of
the old symbol that arrives back at it through $e_{13}$, so we can
then remove the connection between edges $e_{13}$ and $e_{15}$ from
the line graph. Hence the actual line graph followed by the symbols
that enter the knot through edge $e_2$ is shown in Figure 8 (b).

In that graph there is only one cycle, which has length 3, namely
$\{e_{17},e_{14},e_{16},e_{17}\}$. All the transfer functions will
have as denominator the function $\Delta=1+D^3$.

For the transfer function corresponding to edge $e_{15}$, Figure 9
(a) shows that the only forward path is node disjoint with the only
cycle of the graph, hence $F_1(e_2,e_{15})=D$ and
$\Delta_1(e_{15},e_{15})=1+D^3$. Finally $\tau(e_2,e_{15})=\frac{D
\cdot (1+D^3)}{1+D^3}=D$ which is the expectable result.

\begin{figure}[htbp]
\centerline{\includegraphics[width=12cm,keepaspectratio=true,angle=0]{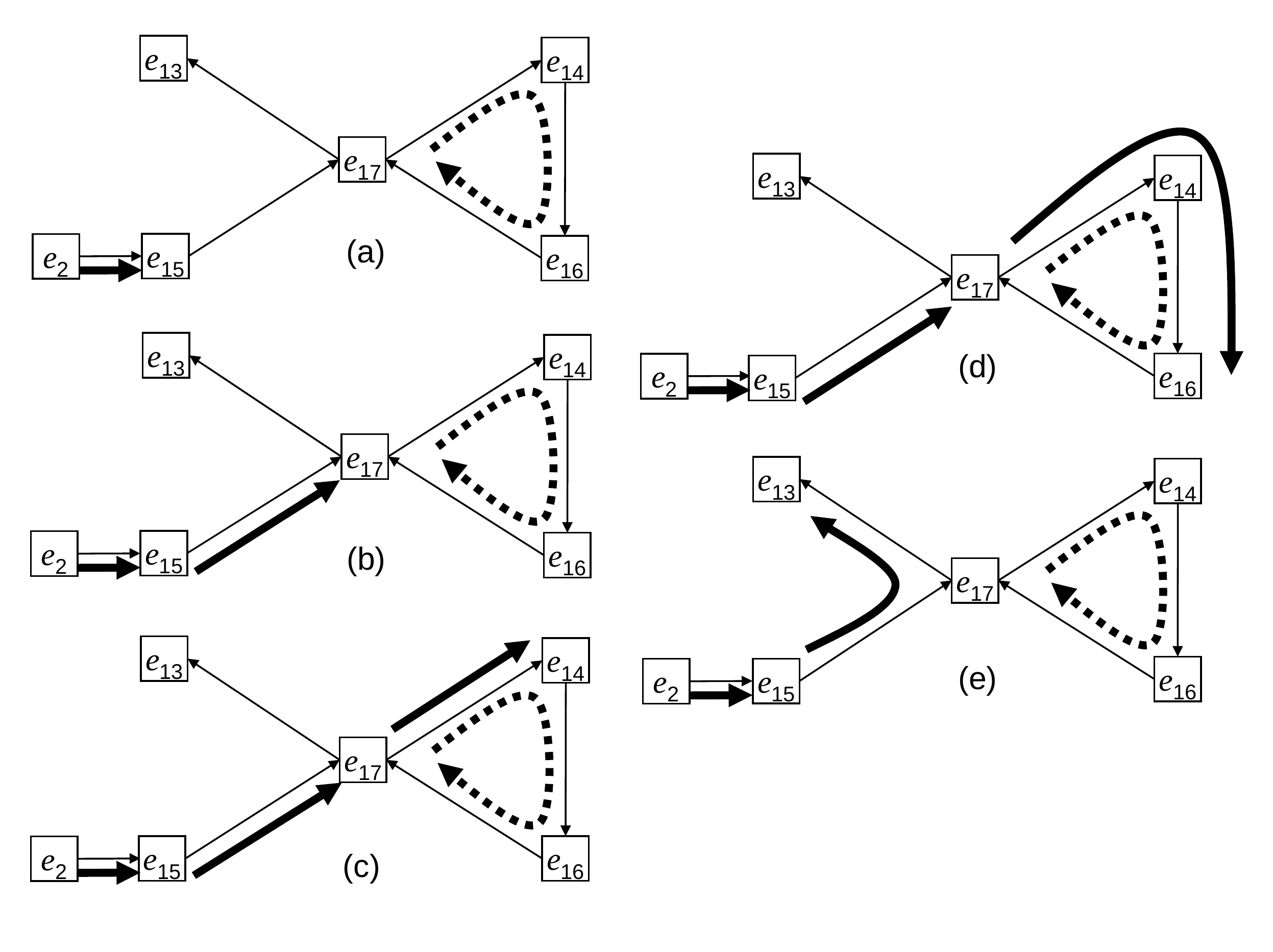}}
\label{Fig_Mason Formula e2 Knot Example4}
\caption{Computation of
Mason's formula from edge $e_2$.}
\end{figure}

Now we compute the transfer function corresponding to edge $e_{17}$.
Figure 9 (b) shows  the only forward path,  which is not node
disjoint with the only cycle of the graph, hence
$F_1(e_2,e_{17})=D^2$ and $\Delta_1(e_2,e_{17})=1$. Finally
$\tau(e_2,e_{17})=\frac{D^2 \cdot 1}{1+D^3}$.

In the same way all the other transfer functions can be computed
(drawings of the corresponding forward paths can be seen in Figures
9 c) to e)).

\[\tau(e_2,e_{14})=\frac{D^3 \cdot 1}{1+D^3}, \tau(e_2,e_{16})=\frac{D^4 \cdot
1}{1+D^3}, \tau(e_2,e_{13})=\frac{D^3 \cdot 1}{1+D^3}\]

Specially interesting is the transfer function corresponding to edge
$e_{13}$ since it gives the function of the old symbol that has to
be removed when passing again through node
$end(e_2)=end(e_{13})=start(e_{15})$.

\item
If we now focus on the circulation in the knot of the symbol carried
by edge $e_5$ we observe (Figure 10) that the flow path will
transport it through edges $e_{13}$ and $e_{15}$ (bold line in the
figure) and node $end(e_{15})$ will send it back to node $end(e_5)$
through edge $e_{17}$ (dashed in the figure), but node $end(e_5)$
will not send that symbol on edge $e_{14}$ and hence the symbol does
not travel along the two cycles in the knot, but only on the cycle
$\{e_{13}, e_{15},e_{17}, e_{13}\}$. The computation of the transfer
functions is then straightforward.

\begin{figure}[htbp]
\centerline{\includegraphics[width=10cm,keepaspectratio=true,angle=0]{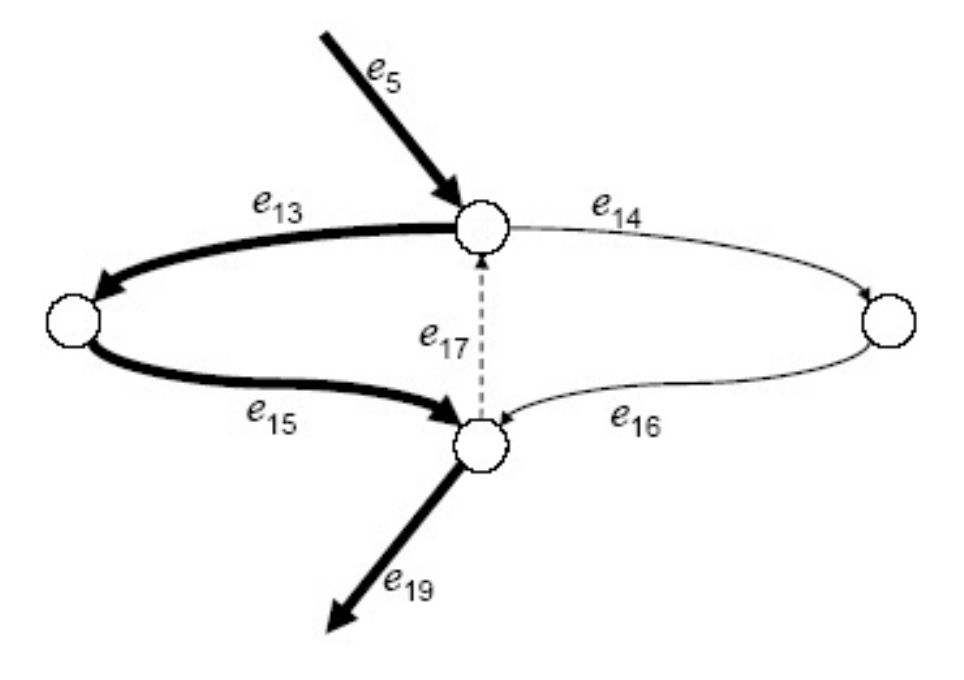}}
\label{Fig_Flow e5 Knot Example4} \caption{Itinerary of symbols
carried by $e_5$ in the knot of Example 4.}
\end{figure}

\[\tau(e_5,e_{13})=D, \tau(e_5,e_{15})=D^2, \tau(e_5,e_{17})=D^3,\tau(e_5,e_{14})=0, \tau(e_5,e_{16})=0 \]

\item
The symbol entering the knot via edge $e_8$ follows a similar
trajectory to that of the one entering via edge $e_5$.

\[\tau(e_8,e_{14})=D, \tau(e_8,e_{16})=D^2, \tau(e_8,e_{17})=D^3,\tau(e_8,e_{13})=0, \tau(e_8,e_{15})=0 \]

\item
Finally, the symbol entering via edge $e_{11}$ follows an itinerary
identical (considering symmetry) to that entering via $e_2$ already
studied.

\[\begin{array}{l}
\tau(e_{11},e_{16})=\frac{D \cdot (1+D^3)}{1+D^3}=D,
\tau(e_{11},e_{17})=\frac{D^2 \cdot 1}{1+D^3},
\\ \\
\tau(e_{11},e_{13})=\frac{D^3 \cdot 1}{1+D^3},
\tau(e_{11},e_{15})=\frac{D^4 \cdot 1}{1+D^3},
\tau(e_{11},e_{14})=\frac{D^3 \cdot 1}{1+D^3}\end{array}\]

\end{itemize}

Now that all the transfer functions have bee computed one can use
formula in Equation (5) to compute the encoding of each edge in the
knot.

We will go in detail with the computation of the encoding of edge
$e_{13}$.

\[\begin{array}{ll}
v_{e_{13}}(x) & =
D^{i_C(e_2)}\tau(e_2,e_{13})v_2(x)+D^{i_C(e_5)}\tau(e_5,e_{13})v_5(x)+
\\
 & + D^{i_C(e_8)}\tau(e_8,e_{13})v_8(x)+
D^{i_C(e_{11})}\tau(e_{11},e_{13})v_{11}(x)\\
 & =D^{i_C(e_2)}\frac{D^3}{1+D^3}a(x-1)+D^{i_C(e_5)}D b(x-1)+D^{i_C(e_8)}0 c(x-1)+
D^{i_C(e_{11})}\frac{D^3}{1+D^3}d(x-1)\end{array}\]

In the same way the encoding of the other edge in the knot can be
computed using Equation (6).

\[\begin{array}{l}
v_{e_{14}}(x) =D^{i_C(e_2)}\frac{D^3}{1+D^3}a(x-1)+D^{i_C(e_5)}0
b(x-1)+D^{i_C(e_8)}D c(x-1)+
D^{i_C(e_{11})}\frac{D^3}{1+D^3}d(x-1)\\ \\
v_{e_{15}}(x) =D^{i_C(e_2)}D a(x-1)+D^{i_C(e_5)}D^2
b(x-1)+D^{i_C(e_8)}0 c(x-1)+
D^{i_C(e_{11})}\frac{D^4}{1+D^3}d(x-1)\\ \\
v_{e_{16}}(x) =D^{i_C(e_2)}\frac{D^4}{1+D^3}a(x-1)+D^{i_C(e_5)}0
b(x-1)+D^{i_C(e_8)}D^2 c(x-1)+ D^{i_C(e_{11})}D d(x-1)\\ \\
v_{e_{17}}(x) =D^{i_C(e_2)}\frac{D^2}{1+D^3}a(x-1)+D^{i_C(e_5)}D^3
b(x-1)+D^{i_C(e_8)}D^3 c(x-1)+
D^{i_C(e_{11})}\frac{D^2}{1+D^3}d(x-1)
\end{array}\]

Now we will show how each edge can compute its encoding based on its
predecessors and using formula in Equation (5). Again we will go in
detail with edge $e_{13}$.

\[\begin{array}{l}
P(e_{13})=\{e_5,e_{17}\},P(e_{13})\cap P(C)=\{e_5\}, P(e_{13})\cap
C_E=\{e_{17}\}\\
P'(e_{13})=\{e_5,e_8, e_{17}\},P'(e_{13})\cap P(C)=\{e_5, e_8\}
\end{array}\]

The direct application of formula in Equation (5) to this case gives
the following

\[\begin{array}{ll}
v_{e_{13}}(x) & = Dv_{e_{17}}(x)+D
D^{i_C(e_5)}v_{e_5}(x)-D\left(D^{i_C(e_5)}\tau(e_5,e_{17})v_{e_5}(x)+D^{i_C(e_8)}\tau(e_8,e_{17})v_{e_8}(x)\right)\\
 & =D\left(D^{i_C(e_2)}\frac{D^2}{1+D^3}a(x-1)+D^{i_C(e_5)}D^3
b(x-1)\right. \\
 & + \left.D^{i_C(e_8)}D^3 c(x-1)+
D^{i_C(e_{11})}\frac{D^2}{1+D^3}d(x-1)\right) \\
 & + D^{i_C(e_5)} D b(x-1) \\
 & - D^{i_C(e_5)} D D^3 b(x-1)-D^{i_C(e_8)} D D^3 c(x-1) \\
 & =D^{i_C(e_2)}\frac{D^3}{1+D^3}a(x-1)+D^{i_C(e_5)}D b(x-1)+D^{i_C(e_8)}0 c(x-1)+
D^{i_C(e_{11})}\frac{D^3}{1+D^3}d(x-1)
\end{array}\]

In a similar way all the other edges can get their encoding using
those of its predecessors and formula in Equation (5).

Next point is determining the unknowns $i_C(e)$ for each edge in $C$
by checking the full rank conditions. To be precise, the full rank
condition must be checked for edge $e_{13}$ in the flow to sink
$t_1$, for edge $e_{14}$ in the flow to sink $t_3$ and for edges
$e_{15}$ and $e_{16}$ in the flow to sink $t_2$. A careful exam of
the corresponding matrices will show that
$i_C(e_2)=i_C(e_5)=i_C(e_8)=i_C(e_{11})=0$ is a valid choice.

Finally, a similar remark to the one made in Example 3 gives us
$v_{e_{18}}(x)=v_{e_{16}}(x)$ and $v_{e_{21}}(x)=v_{e_{15}}(x)$.

Also we have
$v_{e_{19}}(x)=Dv_{e_{15}}(x)$,$v_{e_{20}}(x)=Dv_{e_{16}}(x)$.

The final encoding matrices at the sinks are

\[M_{t_1}=\left(\begin{array}{cccc}
\frac{D^5}{1+D^3} & 0 & 0 & 0 \\
0 & D & 0 & 0 \\
D^3 & 0 & D  & 0 \\
D^2 & 0 & 0 & D
\end{array}\right),
 M_{t_2}=\left(\begin{array}{cccc}
D & D^3 & \frac{D^6}{1+D^3} & 0 \\
0 & D^4 & 0 & 0 \\
0 & 0 & D^4 & 0 \\
0 & \frac{D^6}{1+D^3} & D^3 & D
\end{array}\right),
 M_{t_3}=\left(\begin{array}{cccc}
D & 0 & 0 & D^2\\
0 & D & 0 & D^3 \\
0 & 0 & D & 0 \\
0 & 0 & 0 & \frac{D^5}{1+D^3}
\end{array}\right)\]

As can be seen the elements in the matrices are now rational
functions (typical case after traversing a knot). This however
represents no extra difficulty at the sink. For instance, if we
focus on sink $t_2$, the received symbols at that sink are
$r_{t_2,1}(x)=v_1(x), r_{t_2,2}(x)=v_{19}(x),
r_{t_2,3}(x)=v_{20}(x)$ and $r_{t_2,4}(x)=v_{12}(x)$. Their
expressions in terms of the source symbols are given in each column
of the matrix $M_{t_2}$. In order to retrieve the source symbols
$t_2$ will multiply the received ones by the inverse of matrix
$M_{t_2}$.

\[ \begin{array}{ll}
[a(x),b(x),c(x),d(x)] &
=[r_{t_2,1}(x),r_{t_2,2}(x),r_{t_2,3}(x),r_{t_2,4}(x)]M_{t_2}^{-1}
\\
 & = \left[ r_{t_2,1}(x+1), r_{t_2,1}(x+2)+r_{t_2,2}(x+4)+\frac{D}{1+D^3} r_{t_2,4}(x), \right.\\
  & \;\left. \frac{D}{1+D^3} r_{t_2,1}(x)+r_{t_2,3}(x+4)+r_{t_2,4}(x+2), r_{t_2,4}(x+1)\right]
 \end{array}\]

The way to deal with expressions like $\frac{D}{1+D^3} r_{t_2,1}(x)$
at $t_1$ is to keep in memory three local variables that we will
call $r_{t_2,1,i}(x)$. They will all be initialized as 0, and at
time $x$ one of them will be updated and the other two keep the same
as follows:

\[\begin{array}{lll}
r_{t_2,1,i_x}(x) & = & r_{t_2,1,i_x}(x-1)+r_{t_2,1}(x) \\
r_{t_2,1,i}(x) & = & r_{t_2,1,i}(x-1) \; {\rm for} \; i\in
\{0,1,2\}\setminus \{i_x\} \end{array}\]

where $x=3q_x+i_x$ with $i_x\in\{0,1,2\}$, that is to say, $i_x$ is
the remainder of the integer division of $x$ by 3.

In this way $\frac{D}{1+D^3} r_{t_2,1}(x)=r_{t_2,1,i_{(x-1)}}(x-1)$,
and the receiver $t_2$ does not need to keep an infinite memory,
despite the aspect of the equations received.

\subsection*{Example 5}

We will briefly show here one more example in which a more
complicated network is dealt with.

\begin{figure}[htbp]
\centerline{\includegraphics[width=10cm,keepaspectratio=true,angle=0]{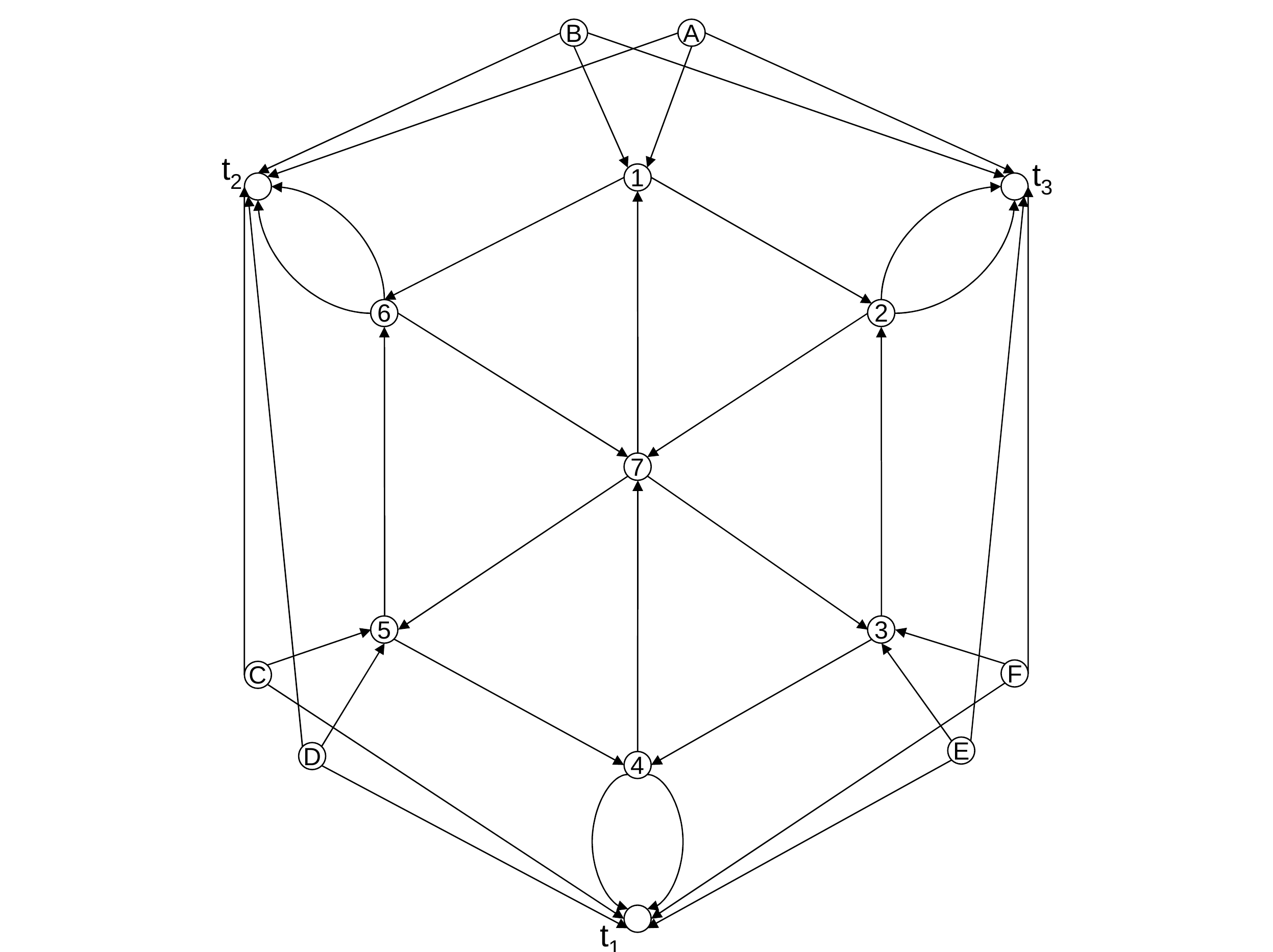}}
\label{Fig_Example5} \caption{The network of Example 5.}
\end{figure}

The network is shown in Figure~11. It has six unit rate sources,
labeled $A,B,\ldots, F$. and three sinks, labeled $t_1,t_2,t_3$.

The other nodes have been labeled 1,\ldots, 7.

\begin{figure}[htbp]
\centerline{\includegraphics[width=10cm,keepaspectratio=true,angle=0]{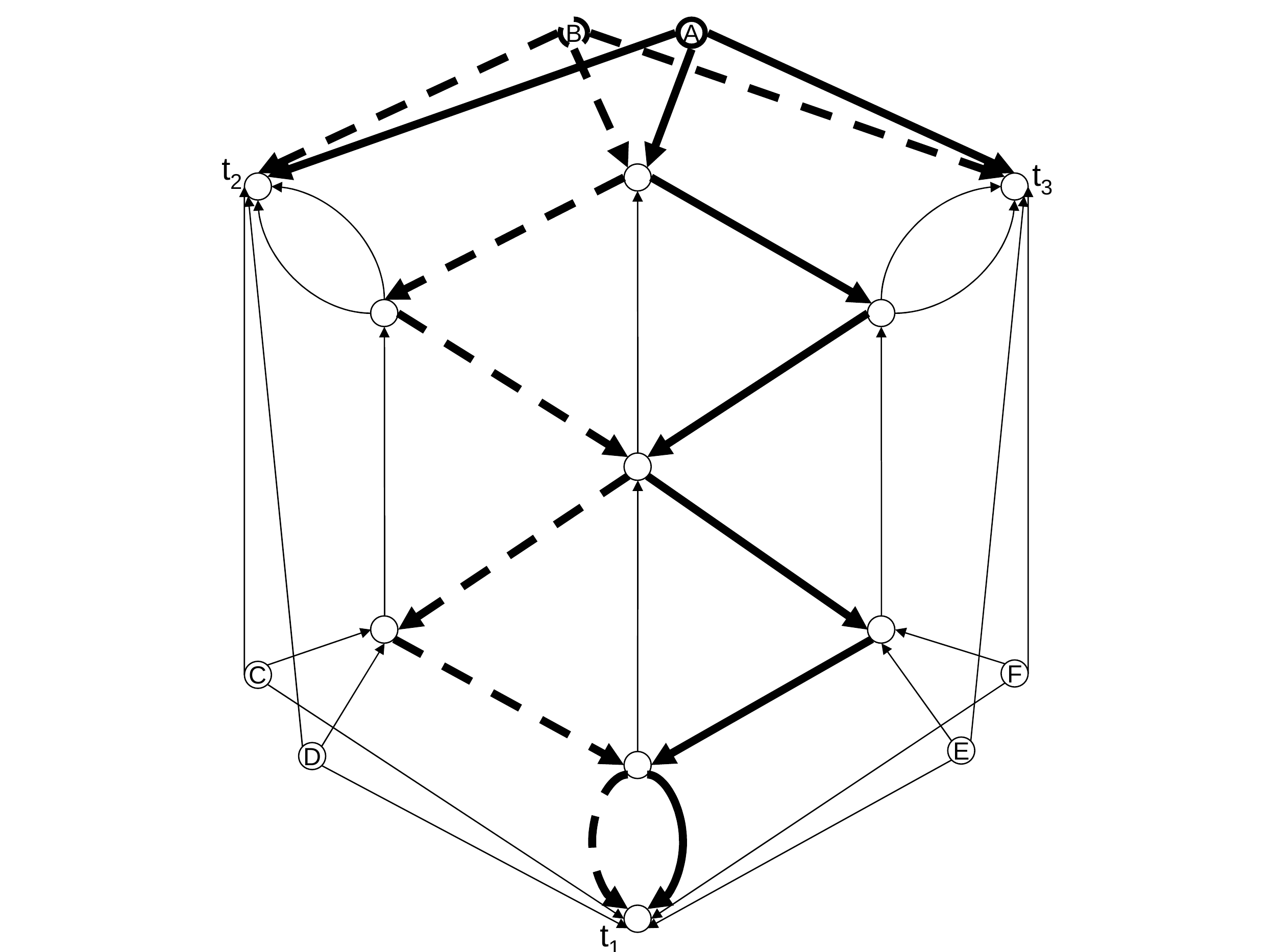}}
\label{Fig_Flow Example5} \caption{Flow paths in the network of
Example 5.}
\end{figure}

The flow path graph for such a network is (essentially) unique. The
flow paths starting in sources $A$ and $B$  are shown in Figure~12
in solid bold and dashed bold lines respectively. The flow paths
from the other sources are the same but with a rotation of 120
degrees to the right or to the left.

\begin{figure}[htbp]
\centerline{\includegraphics[width=13cm,keepaspectratio=true,angle=0]{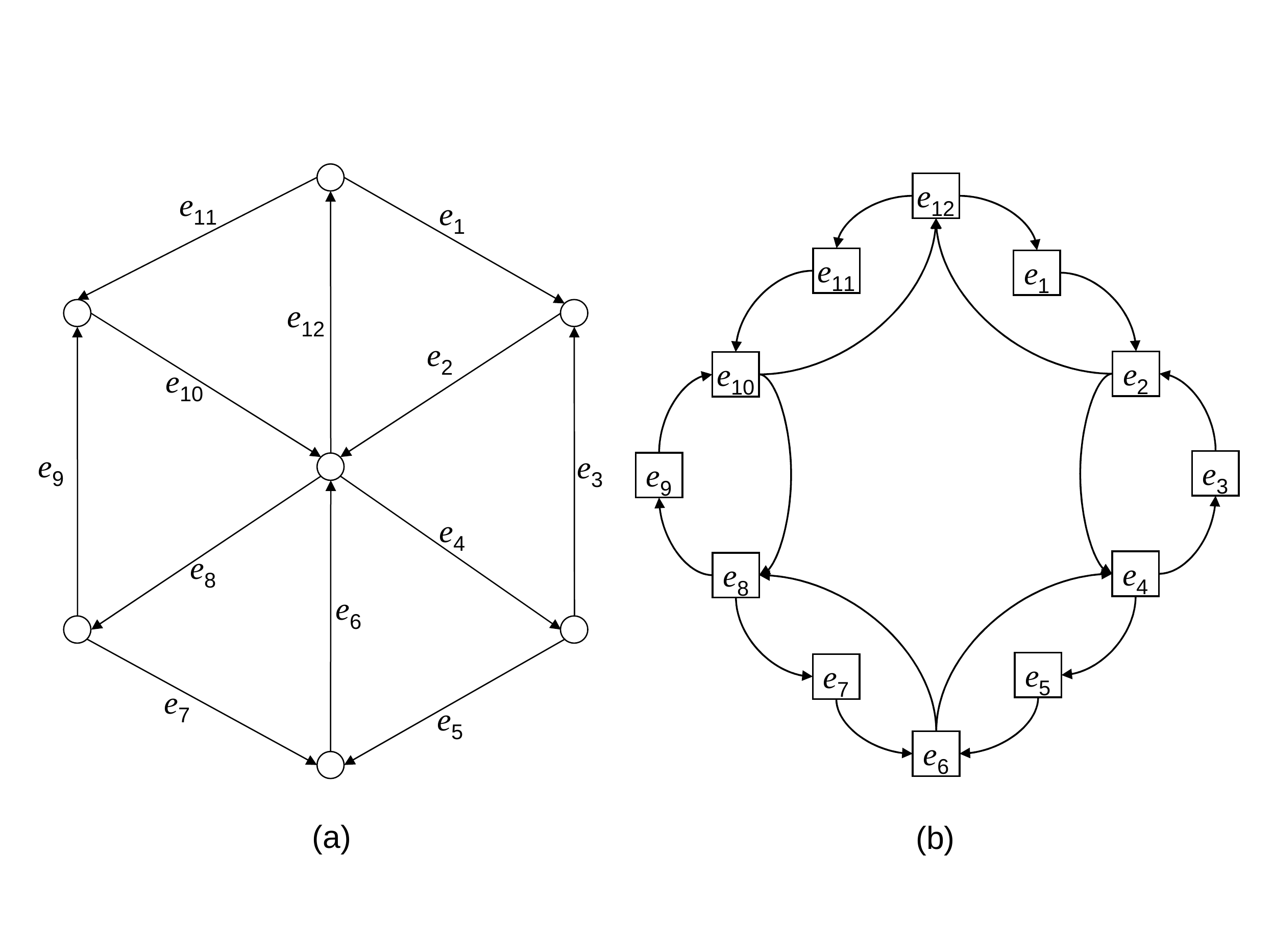}}
\label{Fig_Knot Example5} \caption{The knot and its line graph in
the network of Example 5.}
\end{figure}

This implies that the flow path graph of that network contains a
knot formed by 6 simple cycles, each of length 3. Figure~13 a) shows
the knot while Figure~13 b) shows the corresponding line graph
constructed following the flow paths through the knot.

\begin{figure}[htbp]
\centerline{\includegraphics[width=13cm,keepaspectratio=true,angle=0]{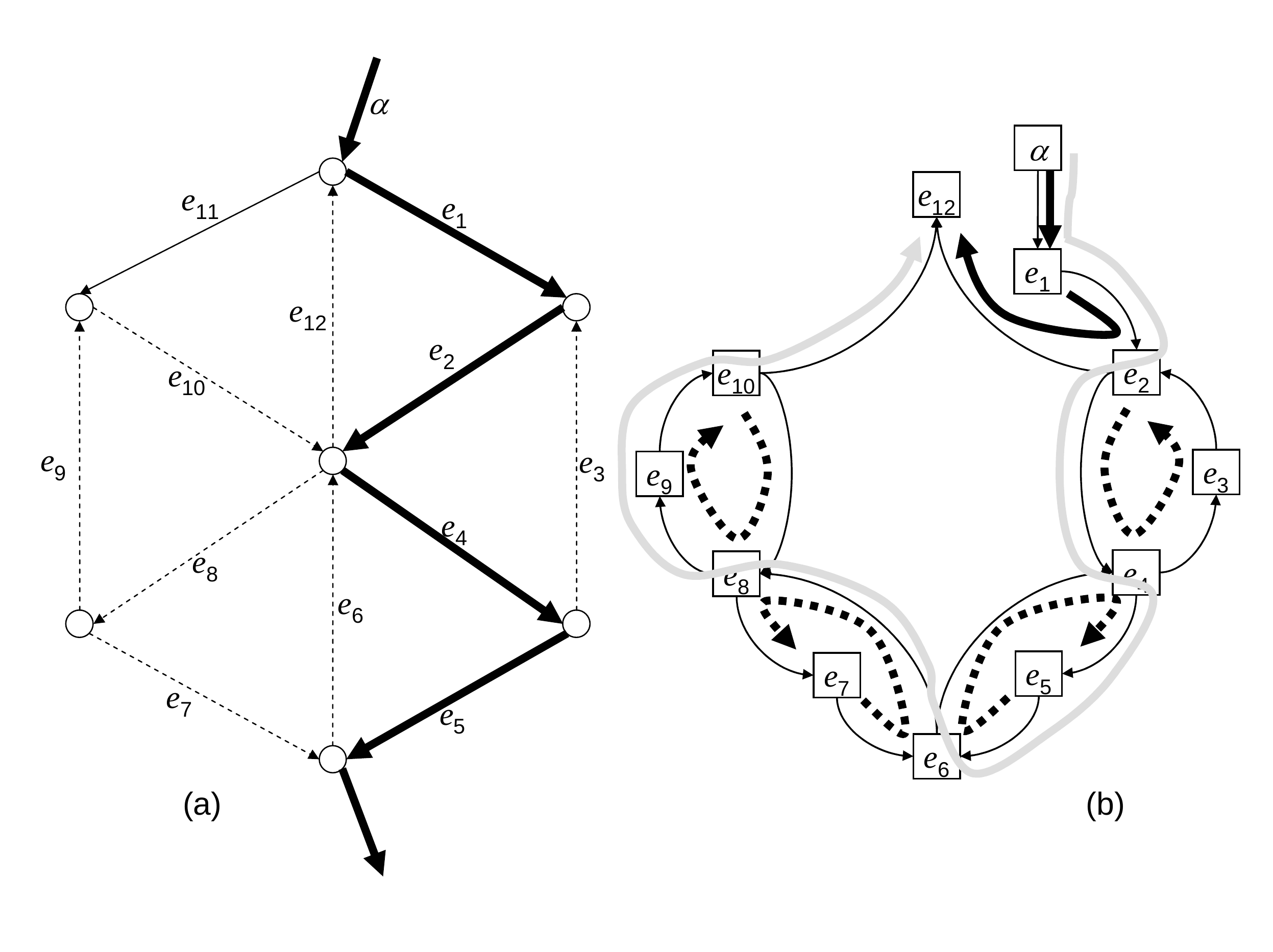}}
\label{Fig_tranfer_Example5} \caption{Itinerary of the symbol from
edge $\alpha$ through the knot in the network of Example 5.}
\end{figure}

We will focus now on the way the symbols are carried by the edge
that connects source $A$ with node $1$ travel through the node.
Mason's formula will be used to compute the transfer functions from
that edge (which is one of the predecessors of the knot) to any
other edge in the knot. Let us call $\alpha$ that edge, that is,
$\alpha=(A,1)\in P(C)$. In Figure~14 a) we show the itinerary of the
symbol from $\alpha$. The bold solid lines are the actual flow path
and the dashed lines are the edges not belonging to the flow path of
the symbol but that will nevertheless carry instances of that symbol
due to the connections in the knot. We can see that the only edge in
which that symbol do not travel is $e_{11}$. Figure~14 b) shows the
corresponding modified line graph that allow us to compute the
transfer function $\tau(\alpha, 12)$. The bold dashed arrows show
the position of the 4 simple cycles in that graph and the two bold
lines (one black and the other grey) show the trajectories of the
two different forward paths from $\alpha$ to $e_{12}$. Direct
application of Mason's formula gives

\[\tau(\alpha,
e_{12})=\frac{D^3(1+3D^3+D^6)+D^9\dot
1}{1+4D^3+3D^6}=\frac{D^3}{1+D^3}\]

The rest of the transfer functions from $\alpha$ to edges in $C$ are

\[\begin{array}{llll}
\tau(\alpha,e_1)=D, & \tau(\alpha,e_2)=D^2+\frac{D^5}{1+D^6},
&\tau(\alpha,e_3)=\frac{D^4}{1+D^6},
 & \tau(\alpha,e_4)=\frac{D^3}{1+D^6},\\
\tau(\alpha,e_5)=\frac{D^4}{1+D^6}, &
\tau(\alpha,e_6)=\frac{D^5}{1+D^3}, &
\tau(\alpha,e_7)=\frac{D^7}{1+D^6}, &
\tau(\alpha,e_8)=\frac{D^6}{1+D^6}, \\
\tau(\alpha,e_9)=\frac{D^7}{1+D^6}, &
\tau(\alpha,e_{10})=\frac{D^8}{1+D^6}, & \tau(\alpha,e_{11})=0,&
\end{array}\]

The transfer functions from the other predecessors of the knot to
the edges in the knot are analogous. In fact, they can be derived
from the ones already computed by simply taking into account the
multiple symmetries that this network presents.

If we denote by $\beta$ the edge that connects source $B$ with node
$1$, $\gamma$ the one that connects $C$ with 5, $\delta$ the one
that connects $D$ with 5, and finally $\epsilon$ and $\phi$ the
edges connecting sources $E$ and $F$ respectively to node 3,
following the procedure of LIFE* we obtain the following global
encoding equation for edge $e_5$

\[v_{e_5}(x)=D^{i_C(\alpha)}\frac{D^5}{1+D^6}a(x)+D^{i_C(\beta)}\frac{D^8}{1+D^6}b(x)+
D^{i_C(\gamma)}\frac{D^8}{1+D^6}c(x)+
D^{i_C(\delta)}\frac{D^5}{1+D^6}d(x)+ D^{i_C(\epsilon)}D^2e(x)\]

Encoding for other edges in the network will be analogous.

The full rank condition will be satisfied when choosing
$e_C(\alpha)=e_C(\beta)=e_C(\gamma)=e_C(\delta)=e_C(\epsilon)=e_C(\phi)=0$.

The local encoding equation of edge $e_5$ in terms of its
predecessors is as follows.

\[v_{e_5}(x)=Dv_{e_4}(x)+DD^{i_C(\epsilon)}v_{\epsilon}(x))-D \left(
D^{i_C(\epsilon)}\tau(\epsilon,
e_4)v_{\epsilon}(x)+D^{i_C(\phi)}\tau(\phi, e_4)v_{\phi}(x)\right)\]

The decoding matrix for sink $t_1$ has the form

\[M_{t_1}^{-1}=\frac{1}{D^6}\left(\begin{array}{cccccc}
1 & D^3 & 0 & 0 & 0 & 0 \\
D^3 & 1 & 0 & 0 & 0 & 0 \\
\frac{D^8}{1+D^6} & \frac{D^{11}}{1+D^6} & D^5 & 0 & 0 & 0 \\
\frac{D^{11}}{1+D^6} & \frac{D^{2}}{1+D^6} & 0 &  D^5 & 0 & 0  \\
D^2+\frac{D^{11}}{1+D^6} & D^5+\frac{D^{8}}{1+D^6} &  0 & 0 & D^5 & 0  \\
\frac{D^8}{1+D^6} & \frac{D^5}{1+D^6} & 0 & 0 & 0 & D^5
\end{array}\right)\]


\end{document}